\documentclass[final,3p,times]{elsarticle}
\usepackage{graphicx}
\usepackage{color}
\usepackage{amssymb}
\usepackage{amsmath}
\usepackage{hyperref}
\usepackage{float}
\usepackage{enumerate}
\usepackage{mathrsfs}
\usepackage[utf8]{inputenc}
\usepackage[T1]{fontenc}
\usepackage{xcolor}
\usepackage{mathrsfs}

\begin{document}
	
	\begin{frontmatter}
		\title{Universal Scaling of the Magnetocaloric Effect in 2D Ising Monolayers and Bilayers}
		\author[vit]{Basit Iqbal}
		\author[vit]{Kingshuk Sarkar \corref{cor1}}
		\cortext[cor1]{Corresponding author}
		\ead{kingshuk.sarkar@vitap.ac.in}
		
		\address[vit]{Department of Physics, School of Advanced Sciences, VIT-AP University, Amaravati, Andhra Pradesh – 522237, India}
		
		\begin{abstract} \underline{}
			We report a comprehensive investigation of the magnetocaloric effect (MCE) in two-dimensional(2D) square, honeycomb, and triangular lattices, considering both monolayer and bilayer configurations by performing Monte Carlo simulations of the ferromagnetic Ising model. These lattice geometries exhibit distinct magnetic responses due to variations in coordination number($r$), which directly influence the number of magnetic exchange interactions($J$). Using Binder cumulant analysis, we find the precise critical temperature $T_c$ of each structure. We find that $T_c$ increases monotonically with the coordination number, being lowest for the honeycomb lattice ($r$=3) and highest for the triangular lattice ($r$=6). In contrast, the magnetic entropy change $(-\Delta S^{}_M)$ exhibits the opposite trend, attaining its maximum value for the honeycomb lattice. Remarkably, when normalized by their peak values, both the $-\Delta S^{}_M$ curves and the field exponent $n$ collapse onto a universal master curve while the temperature is appropriately scaled first, for each lattice across varying magnetic fields, and second, at all six lattice structures at a fixed low field. This demonstrates that the scaling behavior of the MCE is universal, independent of both coordination numbers and layer count. The universality of MCE across different lattice geometries and emerging power law behaviors are rationalized further with the scaling analysis in the critical region. Unlike $-\Delta S^{}_M$, the adiabatic temperature change ($\Delta T_{ad}$) increases with the coordination number, reaching its maximum for the triangular lattice. However, like $-\Delta S^{}_M$, the magnetic Gr\"{u}neisen parameter ($\Gamma_{M}$) decreases with the coordination number, reaching its maximum for the honeycomb lattice. Although the peak values of $\Delta S^{}_M$ decrease with increasing coordination number, the relative cooling power (RCP) and cooling capacity ($q$) are largely unchanged due to a compensating broadening of $\Delta S^{}_M$ curves. The field dependencies of $\Delta S^{}_M$, RCP, and $q$ follow power laws in the conventional field range up to $\sim$ 2.6 Tesla (assuming $J\approx 1 \; meV$). Hysteresis analysis further reveals that lattices with lower coordination numbers--which exhibit larger $\Delta S^{}_M$--show a more rapid reduction in loop width with the per unit increase in temperature. Our results on the scaling behavior of MCE and detailed analysis of magnetocaloric quantities in different monolayer and bilayer lattice geometries provide key insights in designing two-dimensional magnetic refrigerants. 
		\end{abstract}
		
		
	\end{frontmatter}
	
	\section{Introduction}
	
\begin{figure}[H] 
		\centering
		\includegraphics[width=16cm,height=7.5cm,trim=0cm 8cm 0cm 0cm]{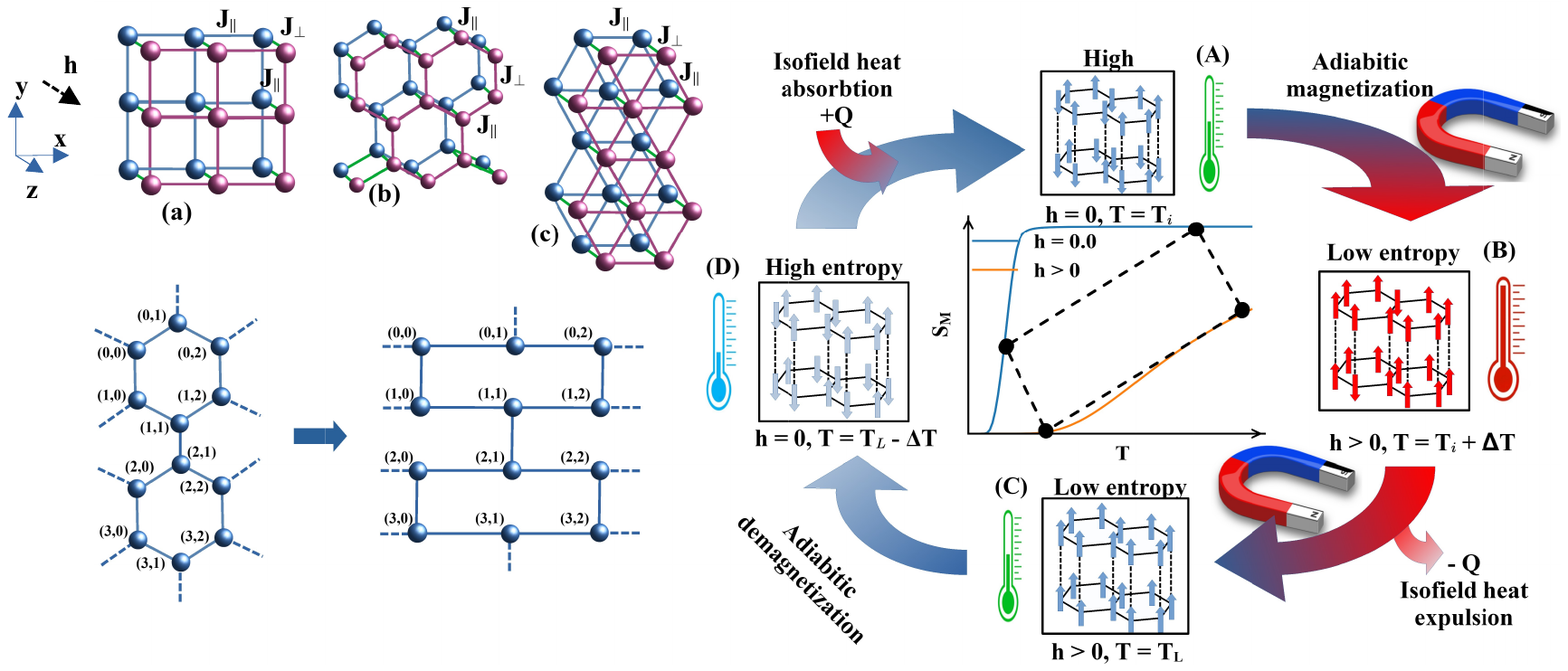}
		\caption{Schematics: (a) square bilayer, (b) honeycomb bilayer, and (c) triangular bilayer. Monolayer lattices can be considered as one layer in these lattices. (d) Schematic representation of the honeycomb lattice and its equivalent brick model. (e) Schematic representation of a MCE cycle as per the Brayton cycle: (A) $\to$(B) adiabatic magnetization, (B) $\to$(C) isofield heat expulsion, (C) $\to$(D) adiabatic demagnetization, and (D) $\to$(A) isofield heat absorption.}
		\label{Figure_1} 
	\end{figure}

	The magnetocaloric effect (MCE) \cite{weiss-1917,tishin-2010} serves as the foundation for magnetic refrigeration, which offers an environmentally less harmful and energy-efficient alternative to conventional gas compression-expansion technology \cite{mcnerney-2025,lucia-2024,gschneidner-2008}. This makes the phenomenon a topic of significantly growing research interest, not only for its applications but also as a powerful probe of critical phenomena and universality in magnetic systems. MCE is an intrinsic property of a magnetic material and describes the thermal response of the material when placed in a magnetic field, resulting in a temperature change due to magnetization and demagnetization. Magnetic entropy ($S_M$) is a thermodynamic quantity that measures the disorder or randomness of the magnetic moments (spins) within a material. When a magnetic material is exposed to an external magnetic field, the magnetic moments of its constituent atoms or ions tend to align along the direction of the field. This process of magnetization reduces the $S_M$ and heats up the system (adiabatically). Conversely, when the field is removed (adiabatically), demagnetization occurs, increasing $S_M$. This can be schematically represented as shown in FIG. \ref{Figure_1}(e). This field-induced change in magnetic entropy, arising from magnetization and demagnetization, is responsible for a crucial physical quantity, the isothermal magnetic entropy change ($\Delta S_M$), which quantifies the magnetocaloric effect \cite{tishin-2010}. Since the change in magnetization is maximum at the phase transition point($T_c$), the change in the magnetic entropy is also maximum near $T_c$. $\Delta S_M$ versus temperature (T), at an applied field, shows a peak near $T_c$, and the area under this $\Delta S_M$ curve between two temperatures $T_1$ and $T_2$ determines the heat exchange, in the refrigeration cycle, known as the refrigeration capacity. So, the refrigeration capacity of a material can be estimated by the peak value of $\Delta S_M(T)$ and its full width at half maximum. MCE is observed in both first-order \cite{pecharsky-1997} and second-order phase transitions \cite{debye-1926,giauque-1927}. As compared to the second-order phase transition, the first-order magnetic phase transition shows larger $\Delta S_M$, but cooling efficiency is lower, as it has larger (thermal, magnetic) hysteresis\cite{franco-2017,gutfleisch-2016}.
	
	Motivated by Widom's scaling relations near the critical region \cite{widom-1965}, the researchers sought to identify scaling laws of the magnetocaloric quantities in magnetic materials undergoing second-order phase transitions (SOPT). In SOPT, the $-\Delta S_M$ increases with field (h). First of all, in 1983, Oesterreicher and Parker, using molecular field models, at and above $T_c$, presented a principle of field-dependence of the magnetic entropy change, $-\Delta S_M \propto h^n$, for SOPT with $n = 2/3$ near $T_c$ \cite{oesterreicher-1984}. Later in 2006, Franco \textit{et al.}\cite{franco-2006} derived a relation of the exponent $n$ with the critical exponents for spontaneous magnetization, \textit{i.e.,} $\beta$, and susceptibility, \textit{i.e.,} $\gamma$, using the Arrott-Noakes equation of state\cite{arrott-1967} for a SOPT. Later in 2011, Lyubina \textit{et al.} \cite{lyubina-2011} derived a more general expression for SOPT that includes higher-order terms in the magnetization. SOPT exhibits a universal behavior in $\Delta S_M$, as proposed by Franco \textit{et al.} based on a phenomenological procedure \cite{franco-2006}. Seminal work by Franco \textit{et al.} \cite{franco-2006} demonstrated that $\Delta S_M$ and $\Delta T_{ad}$ in materials with a SOPT obey a universal scaling law when normalized appropriately. This `master curve' behavior implies that the microscopic details of a material are less important than its universality class. The unanswered question: Does this universality hold across different lattice geometries in 2D? These studies did not show that the universality and scaling relation would persist if the geometry is changed, along with the universality of the exponent in the critical region. More specifically, we ask a fundamental physics question: How do microscopic parameters (geometry) lead to macroscopic universal behavior? $\Gamma_{M}$ is another quantity of of great interest in MCE \cite{liu-2026}. Furthermore, we investigate: does this exhibition of universal behavior based on the phenomenological procedure that exists for $\Delta S_M$ and $\Delta T_{ad}$ exist for $\Gamma_{M}$? The ability to define these curves has significant consequences for material science, as researchers can evaluate the efficiency of new magnetocaloric materials without exhaustive testing. Furthermore, performance can be accurately estimated for extreme temperatures or high magnetic fields that may be beyond the physical capabilities of a specific laboratory's equipment \cite{franco-2006,franco-2008}. 
	
	Guerrero \textit{et al.} presented the existence of the normal and inverse magnetocaloric effect; the existence of various phases; and different orders in the $J_1-J_2$ transverse Ising model with first- and second-nearest-neighbor interactions using the cluster mean-field approach in the presence and absence of the transverse field \cite{guerrero-2020}. Recently, scaling and universal behavior were discussed in relation to cubic structures \cite{alzate-cardona-2019} where the Heisenberg model was considered. Recent first-principles calculations for layered chromium halides $CrX_3$ ($X = F, Cl, Br, I$) and $GdSi_2$ suggest that lower-dimensional materials exhibit greater cooling power than higher-dimensional materials \cite{patra-2024}, making the magnetocaloric study of 2D lattices a crucial area of research. The literature based on experimental and first-principles studies typically reports that, in lower dimensions, materials have the square \cite{schmidt-2007}, honeycomb \cite{zeng-2012, hou-2018}, and triangular \cite{gao-2022,gong-2023} structures. However, this phenomenological universality has been primarily demonstrated for a single material or structure under varying magnetic fields. An open and fundamental question remains: Does this universal scaling hold across different lattice geometries within the same universality class? In other words, if we change the coordination number--a fundamental aspect of the system’s Hamiltonian--do the normalized magnetocaloric response functions still collapse onto a single curve? Two-dimensional (2D) lattices provide an ideal platform to address this question. The square, honeycomb, and triangular lattices exhibit systematically varying coordination numbers ($r$ = 4, 3, and 6, respectively) while belonging to the same 2D Ising universality class. This allows us to isolate the effect of local connectivity on the global scaling behavior of the MCE. Motivated by this, in this work, we study the magnetocaloric effects of these three structures in a single layer and investigate how adding another layer affects their magnetocaloric properties using Monte Carlo simulations of the Ising model. Monte Carlo simulations are important because they explicitly include thermal fluctuations and spatial correlations neglected in mean-field theory, yielding quantitatively reliable results near the exact ones. For example, recent studies used the Monte Carlo simulations based on the classical Ising model to study the magnetocaloric properties of $Gd_5Si_4$, $Tb_5Si_4$\cite{nobrega-2006_2}, $LaMnO_3$\cite{li-2026} and $R_2Fe_{17}$ ($R = Nd$ and $Gd$)\cite{masrour-2019} and found a good agreement with the experimental data showing their usefulness in the understanding the magnetocaloric properties of real materials. The structures we are interested in studying can be schematically represented as shown in the FIG. \ref{Figure_1}(a)-(c). In a separate study, we investigated the magnetocaloric properties of the six lattices and their scaling properties using mean-field approximations \cite{iqbal-2026-mft}. In this work, we go beyond mean-field approximations and employ high-precision Monte Carlo simulations of the ferromagnetic Ising model to investigate the magnetocaloric effect in these three 2D lattice structures, for both monolayer and bilayer cases. We aim to test the universality of the $\Delta S_M$ master curve, the field exponent $n$, $\Delta T_{ad}$, and $\Gamma_{M}$ against changes in lattice geometry and layer number.
	
	After this short introduction, we present the model and formalism in section \ref{M_and_F}--discussing the Hamiltonian in section \ref{SS_Hamiltonian}, Monte-Carlo simulation in section \ref{MCS}, thermodynamic quantities in \ref{TQS} and the magnetocaloric quantities in section \ref{MCE quantites}. The scaling and universality are discussed in section \ref{scaling_and_universality}. After that, in the results and discussion in section \ref{R_D}, we first present the Monte Carlo equilibration and finite size analysis in section of our results in section \ref{FSS_analysis}, followed by finding of precise $T_c$ by the Binder cumulant study in section \ref{TC}, and for the monolayer and bilayer honeycomb, square and triangular we show the results of the thermodynamic quantities in \ref{results}; and magnetocaloric quantities, and scaling and universal behavior in section \ref{MCE_SU}. Additionally, we present the hysteresis study of these structures in section \ref{Hyst}. Furthermore, the discussion and outlook are presented in section \ref{outlook}. Finally, we present the conclusion in section \ref{conc}.
	
	\section{Model and Formalism}
	\label{M_and_F} 
	\subsection{Hamiltonian}
	\label{SS_Hamiltonian}
	To investigate the critical behavior, magnetic properties, and magnetocaloric effects, we consider the ferromagnetic Ising model on two-dimensional (2D) square, honeycomb, and triangular lattices, for both monolayer and bilayer configurations. The Hamiltonian for the monolayer systems is given by: 
	\begin{align}
		H=-J\sum^{}_{\langle i, j\rangle} \sigma^{}_{i}\sigma^{}_{j}-g\mu^{}_Bh\sum^{}_{i}\sigma^{}_i
		\label{Ham_Mono}
	\end{align}
	where $\sigma=\pm 1$ represents the spin at site $i$. In the first term, ${\langle i, j\rangle}$ represents the nearest-neighbor pairs, and the summation of all such pairs is considered. $J$ denotes the exchange coupling constants between such pairs, and for ferromagnetic interaction $J>0$ and antiferromagnetic interaction $J<0$. The second term incorporates the effect of the external magnetic field perpendicular to the two-dimensional plane with field strength $h$; $g$ is the Land\'e $g$-factor, and $\mu^{}_{B}$ is the Bohr magneton. We set $g=1$ and $\mu^{}_{B}=1$, and additionally the Boltzmann constant $k^{}_{B}=1$ for convenience, and therefore the temperatures and magnetic field are evaluated in units of $J/k^{}_{B}$ and $g \mu_B h/ J \equiv h/J$(where for simplification we substitute $g=1$ and $\mu_B=1$) respectively.
	
	For bilayer systems, we include both the intralayer and interlayer interactions, and the Hamiltonian can be written as:
	\begin{align}
		H=-J^{}_{\parallel}\sum_{k=1}^{2} \sum_{\langle i, j\rangle^{}_{k}}^{} \sigma^{}_{i,k} \sigma^{}_{j,k}-J^{}_{\perp}\sum_{i}^{} \sigma^{}_{i,1} \sigma^{}_{i,2}-g\mu^{}_Bh\sum_{k=1}^{2}\sum^{}_{i}\sigma^{}_{i,k}
		\label{Ham_Bi}
	\end{align}
	where $\sigma^{}_{i,k}$ represents the spin at site $i$ in layer $k$, $J^{}_{\parallel}$ is the intralayer exchange coupling between the nearest neighbor pairs in each layer, and $J^{}_{\perp}$ is the interlayer exchange coupling between the two layers. In this study, we consider the bilayer to be symmetric and therefore all the exchange coupling constants are of equal strength \textit{i. e.} \ $J^{}_{\parallel}=J^{}_{\perp}=J$. We also do not consider any crystal field anisotropy term \textit{i. e.} \ $D\sum^{}_{i}\sigma^{2}_{i}$. In spin 1/2 systems, this term effectively becomes a constant; therefore, we choose $D=0$ in our present study. However, this term becomes non-trivial for $\sigma \ge 1$ \textit{i. e.} Blume-Capel model.
	
	In any lattice geometry, the number of nearest neighbors of any lattice site is typically called the coordination number ($r$). Here we consider three different lattice geometries: honeycomb ($r=3$), square ($r=4$), and triangular ($r=6$), both monolayer and bilayer configurations as shown in Fig.\ref{Figure_1}(a)-(c). In bilayer systems, the effective coordination number increases by one due to an additional interlayer interaction. These variations of coordination number provide a systematic way to study the influence of lattice geometry on the magnetocaloric properties.
	\subsection{Monte Carlo simulation}
	\label{MCS}
	For investigating thermodynamic behavior using numerical methods, Monte Carlo(MC) simulations are among the most important and powerful tools for spin systems \cite{landau-2015}. In MC simulations, the Metropolis algorithm is widely used. In this algorithm, a spin is chosen randomly and flipped, and corresponding to that, the change in energy $(\Delta E)$ is calculated in each MC step. The spin flip is always accepted if $\Delta E < 0$, and if $\Delta E > 0$, then the flip is accepted with the probability $e^{-\beta \Delta E}$, where $\beta=\frac{1}{k_BT}$ and $E$ denotes the energy of the given configuration(as discussed below in Eq. \ref{eq_4}). The Metropolis criterion \cite{metropolis-1953} guarantees that the algorithm samples spin configurations with weights proportional to their Boltzmann weights, enabling efficient exploration of the configuration space. When the process is repeated over a sufficient number of Monte Carlo steps, the system evolves towards equilibrium, and thermal averages can be evaluated.
	
	For investigating thermodynamic behavior using numerical methods, Monte Carlo(MC) simulations are among the most important and powerful tools for spin systems \cite{landau-2015}. Exact enumeration of the partition function becomes infeasible for large systems due to the exponential growth in the number of spin configurations as the lattice size increases. MC simulations bypass the problem by sampling representative configurations using the Boltzmann distribution $\approx e^{-\beta E}$. Using stochastic sampling \cite{glauber-1963}, we can calculate thermal averages of various thermodynamic and magnetic quantities (discussed below), which are crucial for the study of phase transitions.
	
	The Metropolis algorithm \cite{metropolis-1953} is selected because it satisfies the detailed balance condition. For Boltzmann probability $\mu( \sigma )$, the detailed balance condition is given by:
	\begin{equation}
		P(\sigma \to \sigma')\mu(\sigma) = P(\sigma' \to \sigma)\mu(\sigma')
		\label{dbc}
	\end{equation}
	where $P(\sigma \to \sigma')$ represents the probability of going from $\sigma$ to $\sigma'$. The detailed balance condition ensures microscopic reversibility of the Markov process \cite{zia-2007}. Detailed balance demands that, in equilibrium, the probability of transitioning from one configuration to another is exactly balanced by the reverse process. This ensures the simulation converges to the correct equilibrium distribution.
	
	In the Wolff algorithm \cite{wolff-1989}, as an alternative to flipping a single spin, the complete cluster of spins pointing in the same direction is flipped. The cluster is built iteratively, starting with a single site called the seed, which is chosen at random on the lattice. After that, the cluster is grown iteratively by checking neighboring sites that are not part of the cluster and adding them if they have the same spin, with probability $e^{-\beta \Delta E}$. Each site in the cluster is checked out once. This way, the cluster continues to grow. In this manner, during a cluster flip, the Markov chain can explore the state space much faster than a single spin-flip process. The Wolff algorithm satisfies the detailed balance condition as in Eq. \ref{dbc} \cite{luijten-2007}. In the Metropolis algorithm, near the transition point, correlations become very strong and extend over large distances, leading to a critical slowing down. The cluster-flipping method introduced by the Wolff algorithm significantly accelerates decorrelation, as indicated by the relation $\tau \approx \xi^{z}$. Here, $\tau$ and $\xi$ represent the correlation time and the correlation length, respectively. $z$ is the critical exponent; the value of $z$ for the Metropolis algorithm is $2.1665 \pm 0.0012$ and $0.25 \pm 001$ for the Wolff algorithm \cite{luijten-2007}. The $z$ in the Monte Carlo simulation of the Markov chain is not universal, and the value of $z$ being very small for the Wolff algorithm indicates that it can overcome critical slowing down much faster than the Metropolis algorithm, near the transition point, in Monte Carlo simulations.
	
	In our study, we focus on the 2D lattices having honeycomb, square, and triangular monolayer structures with 3, 4, and 6 nearest neighbors or coordination numbers($r$), respectively. Furthermore, we extend our investigations to their bilayer counterparts. In our simulations, we chose periodic boundary conditions along the $x$- and $y$-axes in both monolayer and bilayer cases. In the bilayer cases, we performed simulations with periodic and open boundary conditions along the $z$-axis. In the open-boundary condition case, one extra nearest neighbor is added, producing $r=$ 4, 5, and 7 for these three lattices, respectively. In the periodic-boundary condition cases, two extra nearest neighbors are added, producing $r=$ 5, 6, and 8 for these three lattices, respectively. In this way we analysis includes the discussion for $r=$ 3 to 8 of the 2D structures. The 1D atomic chain($r$=2) Ising model does not undergo a thermodynamic phase transition at finite temperatures (\(T > 0\)) because entropy dominates, allowing thermal fluctuations to break ferromagnetic alignments. For the bilayer cases, we present the results of the periodic boundary conditions and discuss the effect for the open-boundary condition cases.
	
	For the generation of the honeycomb and triangular lattices, we incorporate the brick model \cite{wakabayashi-1999, morita-2015}. In simulations of an Ising-like model using Monte Carlo simulations, the lattices remain topologically invariant under lattice transformations under the brick model, in which a lattice is transformed to its brick-type model for the convenience of the simulations \cite{wakabayashi-1999,morita-2015}. Fig. \ref{Figure_1}(d) shows the honeycomb lattice and its equivalent brick model. In the same way, we consider the equivalent brick model for the triangular lattice as well.
	
	For a square lattice, we can take a $N_x \times N_y$ lattice points, where $N_x$ and $N_y$ (with $N_x=N_y$) can be any number and represent the number of lattice points along $x$- and $y$-axis, respectively. As can be estimated from FIG. \ref{Figure_1}(d), to make a honeycomb lattice $N_x$ shall be an odd number $N_x\ge3$ and $N_y$ can any number greater than 2. Similarly, for the triangular lattice, $N_x$ shall be a multiple of 2 and $N_y$ shall be a multiple of 3. For monolayer lattices, we perform the simulations on the lattice sizes of $10 \times 10$, $20 \times 20$, $30 \times 30$, $40 \times 40$, $50 \times 50$, $64 \times 64$ and $128 \times 128$ for the square lattice; $9 \times 10$, $21 \times 20$, $30 \times 30$, $39 \times 40$, $51 \times 50$, $66 \times 64$ and $129 \times 128$ for the honeycomb lattice; and $10 \times 9$, $20 \times 21$, $30 \times 30$, $40 \times 39$, $50 \times 51$, $64 \times 66$ and $128 \times 129$ for the triangular lattice. We call it lattice size, and for the sake of convenience, we represent these six lattice sizes as $\mathscr{L}=10, 20, 40, 50, 64,$ and $128$, respectively, for all three structures, which actually denote the above-mentioned lattice sizes. In the same way, for bilayer lattices, we perform the simulations on the lattice sizes of($N_x \times N_y \times N_z$) $10 \times 10 \times 2$, $20 \times 20 \times 2$, $30 \times 30 \times 2$, $40 \times 40\times 2$, $50 \times 50 \times 2$, $64 \times 64 \times 2$ and $128 \times 128 \times 2$ for the square lattice; $9 \times 10 \times 2$, $15 \times 16 \times 2$, $21 \times 20 \times 2$, $24 \times 24$, $33 \times 32 \times 2$, $66 \times 64 \times 2$ and $129 \times 128 \times 2$ for the honeycomb lattice; and $10 \times 9 \times 2$, $16 \times 15 \times 2$, $20 \times 21 \times 2$, $24 \times 24 \times 2$, $32 \times 33 \times 2$, $64 \times 66 \times 2$ and $128 \times 129 \times 2$ for the triangular lattice. We represent these seven lattice sizes as $\mathscr{L}=10, 16, 20, 24, 32, 64,$ and $128$, respectively.
	
	\subsection{Thermodynamic quantities}
	\label{TQS}
	Different thermodynamic and magnetic quantities are defined and expressed as follows. Magnetization ($M$) gives us an average measure of the orientation of the spins in the lattice. It is the sum of the spins at each site divided by the total number of sites (N). In the Monte Carlo simulation, we compute the system's average magnetization across many spin configurations to ensure the results are statistically reliable. Magnetic susceptibility ($\chi$) is a response function that measures changes in magnetization in response to a small applied magnetic field ($h$), and it quantifies a material's magnetizability. These two quantities $M$ and $\chi$ are related to each other by the fluctuation-dissipation theorem, which says the system response to an external magnetic field is directly proportional to the spontaneous fluctuation of magnetization at equilibrium. The fluctuation of magnetization is quantified by the variance of magnetization, which is the second-order cumulant ($\kappa^{}_2=\langle M^2 \rangle -\langle M \rangle^2 $) of magnetization. The first-order cumulant $\kappa^{}_1=\langle M \rangle$ is magnetization itself. Now, there can be the third-order cumulant ($\kappa^{}_3$) of magnetization, which quantifies the skewness of the magnetization and is relevant for first-order phase transitions, but as we focus on the second-order phase transition here, we do not consider it here. The fourth-order cumulant of magnetization ($\kappa^{}_4$) is very crucial for our discussion of a second-order phase transition. The fourth-order cumulant of the magnetization is used to construct and to define the Binder cumulant ($U_L$). It can be understood from the kurtosis of the magnetization distribution, defined as $\frac{\langle(M-\langle M \rangle )^4\rangle}{(\langle(M-\langle M \rangle )^2\rangle)^2}$. For the Ising model we consider here, the kurtosis reduces to $\frac{\langle M \rangle^4}{\langle M^2 \rangle^2}$. For the disordered state above $T_c$, the magnetization distribution is Gaussian and the kurtosis turns out to be 3. We define the excess kurtosis by subtracting the Gaussian value; therefore, the expression of excess kurtosis reduces to $\frac{\langle M \rangle^4}{\langle M^2 \rangle^2}-3$. Therefore, in the paramagnetic state $T>T_c$, this excess kurtosis reduces to zero, and in the ferromagnetic state $T<T_c$ at much lower temperature, while the magnetization distribution becomes two Dirac delta functions, the excess Kurtosis value approaches -2. For convenience, the Binder cumulant ($U_L$) is defined as "minus one-third times the excess kurtosis" such that in the completely ordered ferromagnetic state $U_L \to 2/3$ and completely disordered paramagnetic state $U_L \to 0$, and the general expression becomes \ref{eq_UL}. At $T=T_c$, the magnetization distribution takes a universal non-Gaussian shape, and the system exhibits scale invariance; therefore, the Binder cumulant approaches a universal value $U^{*}_L$ for a second-order phase transition. For 1st-order phase transitions, such a universal value does not exist. For Ising and Ising-like models, the Binder cumulant, evaluated as a function of temperature for different lattice sizes, is commonly used to determine the precise transition temperature in numerical simulations. The intersection points of $U_L$ versus T plots for different lattice sizes (L) indicate the system's precise transition point. Mathematically, these quantities are written as:
	
	\begin{align}
		& M(T, h)=\Big\langle \frac{1}{N} \sum_{i} \sigma_{i} \Big\rangle \label{eq_2} \\
		& \chi (T, h)=\frac{\partial \langle M \rangle}{\partial h}\Bigg|^{}_{h
			\to 0}=\frac{N}{k_{B} T}\left(\left\langle M^{2}\right\rangle-\langle M\rangle^{2}\right) \label{eq_3} \\
		&U_L(T) = 1 - \frac{\langle M^4 \rangle_L}{ 3 \langle M^2 \rangle^2_L} \label{eq_UL}
	\end{align}
	The angular brackets in the above equation mean the thermodynamic averages. The energy (E) of the Ising model represents the average strength of the system's magnetic interactions. Magnetic specific heat ($C_M$) quantifies the amount of energy a material can absorb per unit temperature increase. Magnetic entropy ($S_M$) gives us the degree of disorder in the system. For the Ising model, entropy can be inferred from specific heat, using the thermodynamic relationship given in Eq. \ref{eq_6}. Mathematically, the above quantities can be written as follows:
	
	\begin{align}
		& E(T, h)=\frac{1}{N} \langle H \rangle \label{eq_4} \\
		&C_M(T, h)
		=\frac{\partial E}{\partial T}=\frac{1}{N k_{B} T^{2}}\left(\left\langle E^{2}\right\rangle-\langle E\rangle^{2}\right) \label{eq_5} \\
		&S_M(T, h)=\int_{0}^{T} \frac{C_M}{T} dT \label{eq_6}
	\end{align}
	The total entropy $(S_T)$ of the system, in addition to magnetic entropy ($S_M$) also has contributions from lattice entropy ($S_l$) and electronic entropy($S_e$) and therefore $S_T = S_M + S_l(T) + S_e(T)$ \cite{nobrega-2006}. 
	\subsubsection{Magnetocaloric quantities}
	\label{MCE quantites}
	The rate of change in magnetization with temperature $\left(\frac{\partial \mathrm{M}}{\partial \mathrm{T}}\right)_h$ gives the temperature dependence of the order parameter (M) of the Ising model. This requires calculating the temperature derivative of magnetization $\left(\frac{\partial \mathrm{M}}{\partial \mathrm{T}}\right)_h$ at different fields. The relation between the $\left( \frac{\partial M \left(T, h \right)}{\partial T}\right)_h $ to the entropy as shown in Eq. \ref{eq_8} comes from Maxwell's relations $\left( \frac{\partial S_M \left(T, h \right)}{\partial h}\right)_T = \left( \frac{\partial M \left(T, h \right)}{\partial T}\right)_h$. The Monte Carlo method allows this to be estimated by taking small temperature steps and analyzing how $\left(\frac{\partial \mathrm{M}}{\partial \mathrm{T}}\right)_h$ changes. The relative cooling power (RCP) of the lattice between two points is estimated by the integration of $\Delta S_M$ in a particular temperature range. Adiabatic temperature change ($\Delta T_{ad}$) is another important part of the measurement of the magnetocaloric effect that measures the change in temperature as the magnetic field strength is increased adiabatically. The equation implies that for $\Delta T_{ad}$ to be high for a material, the $C_M$ should be small and $\Delta S_{M}(T, h)$ or $\frac{\partial M}{\partial T}_h$ should be large. Mathematically, the relations are as follows:
	\begin{align}
		& \left(\frac{\partial M}{\partial T}\right)_{h}=\frac{1}{k_B T^{2}}(\langle M E\rangle-\langle M\rangle\langle E\rangle) \label{eq_7} \\
		& \Delta S_{M} = S_{M}(T, h)-S_{M}(T, 0) =\int_{0}^{h}\left(\frac{\partial M}{\partial T}\right)_{h^{\prime}} d h^{\prime} \label{eq_8}\\
		& RCP=\int_{T_{1}}^{T_{2}} \Delta S_{M}(T) dT, \quad q = \Delta S_M^{MAX} \delta_{FWHM} \label{eq_9} \\
		& \Delta T_{ad} = - \int_{0}^{h}\frac{T}{C_M(h^{\prime},T)}\left(\frac{\partial M}{\partial T}\right)_{h^{\prime}} d h^{\prime} \label{eq_10} \\
		& \Gamma_M = -\frac{\left(\frac{\partial M}{\partial T}\right)_{h}}{C_{M}|_h} = -\frac{\left(\frac{\partial S_M}{\partial h}\right)_{T}}{C_{M}|_h} = -\frac{1}{T}\frac{\left(\partial S/\partial h\right)_{T}}{\left(\partial S/\partial T\right)_{h}} \label{eq_11} = \left. -\frac{1}{T}\frac{\partial T}{\partial h} \right|_{S_M} \\
	\end{align}
	
	For numerical simulation, using Eq. \ref{eq_5}, \ref{eq_7} and \ref{eq_11}, the $\Gamma_{M}$ can also be defined as:
	\begin{align}
		& \Gamma_{M} = -\frac{\left(\frac{\partial M}{\partial T}\right)_{h}}{C_{M}|_h} = - \frac{1}{N}\frac{\langle M E \rangle - \langle M \rangle \langle E \rangle}{\langle E^2 \rangle - \langle E \rangle^2} 
		\label{eq_11b}
	\end{align}
	Here, $\delta_{FWHM}$ is the full width at half maximum of the temperature range around the peak since the lattice and electronic entropy, as given by the Debye and Sommerfeld model, are independent of the field. So, the total entropy change ($\Delta S_T$) here occurs because of the field-dependent part, \textit{i.e.,} the magnetic part. So, it will be a logical inference to say $\Delta S_T = \Delta S_{M}$. The RCP and $q$ are used interchangeably in the literature; we refer to $q$ as the cooling capacity \cite{de-castro-2025,li-2023,yang-2022}. As one can see from Eq. \ref{eq_5}, \ref{eq_6}, and \ref{eq_10}, the contributions to the total heat capacity from the lattice and conduction electron subsystems will be added in the denominator. They will cause a reduction in MCE or $\Delta T_{ad}$. These contributions are material-specific and are excluded here. 
	
	The Gr\"{u}neisen parameter \cite{gruneisen-1908,gruneisen-1912} $\Gamma = -\left(\frac{\partial \ln T}{\partial \ln V}\right)_S = - \frac{V}{\omega_0}\frac{\partial \omega_0}{\partial V}$ (V and $\omega_0$ refer to the volume and phonon frequency, respectively) initially put forward to investigate how the volume changes (arising mainly due to the structural phase transitions) in a crystal lattice effects the vibrational frequencies has become a point of interest in all caloric studies. The magnetic Gr\"{u}neisen parameter ($\Gamma_M$) \cite{yu-2020,zhu-2003} is the magnetic analogue of the conventional $\Gamma$. For the classical Ising magnets, the $\Gamma_M $ plays the same role as the $\Gamma(V, \omega_0)$ does for a structural phase transition: it quantifies the adiabatic temperature change upon varying the magnetic field (instead of pressure/volume). In the magnetocaloric studies, $\Gamma_M$ characterizes how a change in the magnetic field translates into a change in temperature. While the Ising model has a well-defined phase transition at $T_c$ in the absence of a field, once we apply a magnetic field ($h > 0$), the sharp symmetry-breaking transition disappears into a crossover. At zero field $h=0$ and $T \to T_c$, $\Gamma_M$, being second-order derivatives with respect to free energy \cite{yu-2020}, diverges at $T_c$ due to critical fluctuations. For small $h>0$, the divergence is rounded into a finite peak whose height decreases with the increase in the value of h. $\Delta T_{ad}$ can be written in terms of $\Gamma_{M}$ as $\Delta T_{ad} = -\int_0^h T \Gamma_M dh$. For the classical ferromagnetic Ising model, since $\frac{\partial M}{\partial T}<0$ and $C_M>0$ always, therefore, $\Gamma_M > 0$ everywhere, confirming a conventional (cooling) MCE upon adiabatic demagnetization.
	
	The line of maxima of a MCE quantity such as $\Delta S_M$ ($\Delta S_M^{Max}$) with respect to $T$ (similarly $\Gamma_{M}$ and $\Delta T_{ad}$) at different fields in the phase diagram shows the streak where the magnetic fluctuations (like heat capacity and susceptibility, \textit{etc.}) are at their peak known as the supercritical region. In this region, physical properties don't change abruptly but show large fluctuations. For $h>0$, the system is in the supercritical region where no true phase transition occurs; instead, $\Gamma_M(T)$ exhibits a field‑tuneable peak characteristic of a crossover.
	
	\section{Scaling and universality}
	\label{scaling_and_universality}
	One commonly used formalism for scaling analysis, proposed by Eugene Stanley \cite{stanley-1999}, has been applied to obtain the scaling behavior of $\Delta S_M$ in the critical region of second-order phase transitions \cite{smith-2014}. Although the free energy $(F(T, h))$ consists of both a singular part and a non singular part $(F(T, h) = F_{sing.}(t, h)+F_{non-sing.}(T))$, where $t = (T-T_c)/T_c$, this scaling theory applies only to the singular part of the free energy and is applicable only asymptotically close to the $T_c$. Here onwards, we consider only the singular part of $F(T, h)$. Therefore, for simplification $F(T, h)$ $\equiv$ $F(T, h)_{sing.}$. In the second-order phase transition, $F(T, h)$ near $T_c$ and $h = 0$, \textit{i. e.} $F(t, 0)$ will not be an analytic function of $T$, which means that all derivatives will not exist. The fundamental premise of this theory is that the singular component of the free energy, $F(t, 0)$, behaves as a generalized homogeneous function near the critical threshold. Mathematically, for any positive scaling factor $\lambda$, there exist constants $a_T$ and $a_h$ such that:
	\begin{equation}
		F(\lambda^{a_T} t, \lambda^{a_h} h) = \lambda F(t, h)
	\end{equation}
	The relation between the magnetization and free energy is given by the relation $M = -\left( \frac{\partial F}{\partial h} \right)_T$. So, after differentiating 
	$F$ with respect to $h$ and relating $a_T$ and $a_h$ with the critical exponents, the scaling relation for $M$ is given by \cite{smith-2014}:
	\begin{equation}
		M(t, h) = h^{1/\delta} f_M\left( t h^{-1/\Delta} \right)
		\label{Eq_M_Scaling}
	\end{equation}
	where $f_M$ is the scaling function for $M$ and $\Delta$ is the thermodynamic critical scaling exponent. 
	
	The negative rate of change in the $F(T)|_h$ with respect to temperature measures the entropy \textit{i. e.} -$(\partial F / \partial T)_h = S$ which can be written in terms of the critical exponents as\cite{smith-2014}: 
	\begin{equation}
		S(t, h) = h^{1 - \Delta^{-1} + \delta^{-1}} f_S\left( t h^{-1/\Delta} \right) = h^{(1-\alpha)/\Delta} f_S\left( t h^{-1/\Delta} \right)
		\label{S_Scaling}
	\end{equation}
	Here, $f_S$ is the scaling function for $S$; and the thermodynamic critical scaling exponents $\alpha$($C \sim t^{-\alpha}$), $\delta$($M \sim h^{1/\delta}$) and $\Delta$ can be related with the Griffith's equality $\alpha + \Delta(1 + \delta^{-1}) = 2.$ 
	
	The change in entropy is defined as the difference between the entropy values at zero field and at some finite field at a particular temperature. In magnetocaloric studies, the initial field is usually taken to be zero. So, the change in entropy can be written as \cite{franco-2006,smith-2014}
	\begin{equation}
		\Delta S_{M} = S(T, h) - S(T, 0) = h^{(1-\alpha)/\Delta} \tilde{f}_S\left( t h^{-1/\Delta} \right)
		\label{Delta_S_scaling}
	\end{equation}
	where $\tilde{f}_S$ is another scaling function with a new scaling variable $x= t h^{-1/\Delta}$. As is apparent, being just the difference in the entropy, the scaling of $\Delta S_M$ is similar to the scaling of S. In general, Eq. \ref{Delta_S_scaling} implies that $\Delta S_M$ curves of different $h$ values if plotted $\Delta S_M h^{-(1-\alpha)/\Delta}$ versus $t h^{-1/\Delta}$, the curves will collapse on top of each other. Nevertheless, the $\tilde{f}_S$ for different latices can be similar but usually will not be the same, even if their critical exponents are the same \cite{franco-2006,smith-2014}. As discussed earlier, only the magnetic entropy changes with the field. So, we take $\Delta S = \Delta S_M$.
	
	The quantity $\Delta S_M$ is maximum near the transition temperature and is represented by $\Delta S_M^{Max}$ \textit{i.e.,} $\Delta S_M(t\approx 0, h) \approx |\Delta S_M|_{T=T_c} \approx \Delta S_M^{Max}$. Eq. \ref{Delta_S_scaling} shows that field dependence of $\Delta S_M$ follows a power law with the field given by \cite{franco-2006}
	
	\begin{equation}
		|\Delta S_M|_{T=T_c} \propto h^n
		\label{h^n}
	\end{equation}
	where the exponent $n$ is a temperature- and field-dependent parameter for which the general expression can be obtained from Eq.\ref{Delta_S_scaling} \cite{franco-2006,smith-2014}
	
	\begin{equation}
		n = \frac{d (\ln|\Delta S_M|)}{d \ln h} = \frac{h}{\Delta S_M}\left(\frac{\partial M}{\partial T}\right) = \frac{1-\alpha}{\Delta} - \frac{1}{\Delta} \left. \frac{d \ln \tilde{f}_S(x)}{d \ln x} \right|_{x=t h^{-1 / \Delta}}
		\label{n_T}
	\end{equation}
	
	The above equation, in terms of the critical exponents, predicts that when plotted as a function of $t h^{-1 / \Delta}$ in the critical region, the exponent $n$ for different $h$ values will collapse onto a single curve. The value of $n$ at temperatures below $T_c$ approaches 1, indicating that the field independence of magnetization curves, although it depends on temperature. At temperatures above $T_c$, the value of $n$ tends to 2 as an outcome of the Curie-Weiss law \cite{franco-2006} where in the paramagnetic state, $\Delta S_M$ is proportional to $h^2$. At T = $T_c$ the value of $n$ is dependent on the thermodynamic critical exponents $\beta$($M \sim t^{\beta})$ and $\gamma$ ($\chi \sim t^{-\gamma}$) or $\delta$ given by the relation: 
	
	\begin{equation}
		n(T_c) = 1 + \frac{\beta - 1}{\beta + \gamma} = 1 + \frac{1}{\delta}\left(1-\frac{1}{\beta}\right)
		\label{n_T_c}
	\end{equation}
	Eq. \ref{h^n} tells us how fast the peak of $\Delta S_M$ grows, but to determine the $q$ of a system, as shown by Eq. \ref{eq_9}, the width of $\Delta S_M$ curves also plays a crucial role. Following Eq. \ref{Delta_S_scaling} and Eq. \ref{eq_9}, a power law can be derived for q. Since, $\delta_{\mathrm{FWHM}}$ corresponds to width in temperature $\delta_{\mathrm{FWHM}}=T_{r_2}-T_{r_1}$, therefore, $ \delta_{\mathrm{FWHM}} \propto h^{1/\Delta}$ and, as discussed earlier, $\Delta S_M^{Max} \propto h^{(1-\alpha)/\Delta}$. Thus the field dependence of $q$ in the critical region reduces to
	\begin{equation}
		q \propto h^{(2-\alpha)/\Delta} = h^{1 + 1/\delta}
		\label{q_scaling}
	\end{equation}
	This relation has been applied to the experimental data with good agreement in the field range of $1\times 10^4 Oe$ to $4\times10^4 Oe$ \cite{liu-2019}. The RCP also follows the same power law, but in our study, we found that $q$ follows the power law better than the RCP.
	
	The specific heat, in terms of definition, is defined as $C = T(\partial S/\partial T)_h$. So, from Eq. \ref{S_Scaling} $C(t,h) = h^{-\alpha/\Delta} f_C(th^{-1/\Delta})$ and $(\partial S/\partial h)_T=h^{(1-\alpha)/\Delta - 1} f_1(th^{-1/\Delta})$ where $f_C$ and $f_1$ are the scaling functions. Therefore, the scaling function using Eq. \ref{eq_10} for $\Delta T_{ad}$ is given by:
	\begin{equation}
		\Delta T_{ad}(t,h) = h^{1/\Delta}f_T\left(th^{-1/\Delta}\right)
		\label{deltaT_scaling}
	\end{equation}
	Here, $f_T$ is the scaling function, and the constants can be related by $\Delta=\beta\delta$. The scaling theory predicts that the amplitude of $\Delta T_{ad}$ grows with the field as ($h^{1/\Delta}$).
	
	From the scaling relations of M and $C_M$ as given before, it is possible to obtain the scaling relation for $\Gamma_{M}$, which can be written as
	\begin{equation}
		\Gamma_{M}(t,h)=h^{\frac{1}{\beta\delta}-1}f_{\Gamma_{M}} \left(th^{-1/\Delta}\right)
	\end{equation}
	
	where $f_{\Gamma_{M}}$ is scaling function for the $\Gamma_{M}$. In this way, the scaling theory predicts that the growth in $\Gamma_{M}$ with field is proportional to $\frac{1}{\beta\delta}-1$ and if we plot $\Gamma_{M}(t,h)h^{-(\frac{1}{\beta\delta}-1)}$ with respect to $th^{-1/\Delta}$, the curves will collapse to a single master curve that confirms the universality of $\Gamma_{M}$.
	
	Now, going back to the scaling and universality of $\Delta S_M$ as proposed by Franco \textit{et al.} \cite{franco-2006}. Following a phenomenological approach and using Arrott-Noakes equation \cite{arrott-1967}, they proposed to normalize $\Delta S_M$ with $\Delta S_M^{Max}$, \textit{i.e.,} $\Delta S_M/\Delta S_M^{Max}$ and chose reference temperatures $T_{r1}$ and $T_{r2}$, where $\Delta S_M =\frac{1}{2}$ $\Delta S_M^{Max}$ below and above $T_c$, respectively. However, these reference temperatures can be chosen such that $\Delta S_M^{Max}$ is reduced to a fraction (different than 1/2) of its peak value. They also proposed to rescale the temperature with reference to $T_c$ as follows.
	
	\begin{equation}
		\theta = 
		\begin{cases} 
			-\frac{T - T_c}{T_{r_1} - T_c} & T \leq T_c \\
			\frac{T - T_c}{T_{r_2} - T_c} & T > T_c 
		\end{cases}
		\label{eq_theta}
	\end{equation}
	We define $t^{}_{1/2}=\frac{T_{r_1} - T_c}{T_c}$, which follows $x^{}_{1/2}=t^{}_{1/2}h^{-1/\Delta}$. Similarly, they have proposed scaling and collapse of $\Delta T_{ad}$ with respect to $\theta$. We found that the quantity $\Gamma_M$ can also be scaled, and a collapse of the curves can be observed. Using this relation, we identify $\theta=t \times t^{-1}_{1/2}$. As $t^{-1}_{1/2} \propto h^{-1/\Delta}$, we argue that the rescaled temperature $\theta \propto x \propto th^{-1/\Delta}$. 
	
	The variation of $\Delta S_M/\Delta S_M^{Max}$, the exponent $n$, $\Delta T_{ad}/\Delta T_{ad}^{Max}$, and $\Gamma_M/\Gamma_M^{Max}$, with respect to $\theta$ curves for different field values collapses to a single master curve for low to moderate fields \cite{smith-2014}. This demonstrates the universal behavior of these magnetocaloric quantities. The collapse of these magnetocaloric quantities as discussed above with respect to $\theta$ and as per their scaling laws is the consequence of the scaling behavior \cite{franco-2006,franco-2008,franco-2010}. The collapse of the curves is observed for the materials exhibiting second-order phase transition and is not exhibited by the first-order phase transition \cite{smith-2014,bonilla-2010}. So, it is used to establish the true order of the phase transition.
	
	\section{Results}
	\label{R_D}
	
	\subsection{Monte Carlo equilibration and finite size analysis}
	\label{FSS_analysis}
	
\begin{figure}[H]
		\centering
		\includegraphics[width=16cm,height=5.5cm]{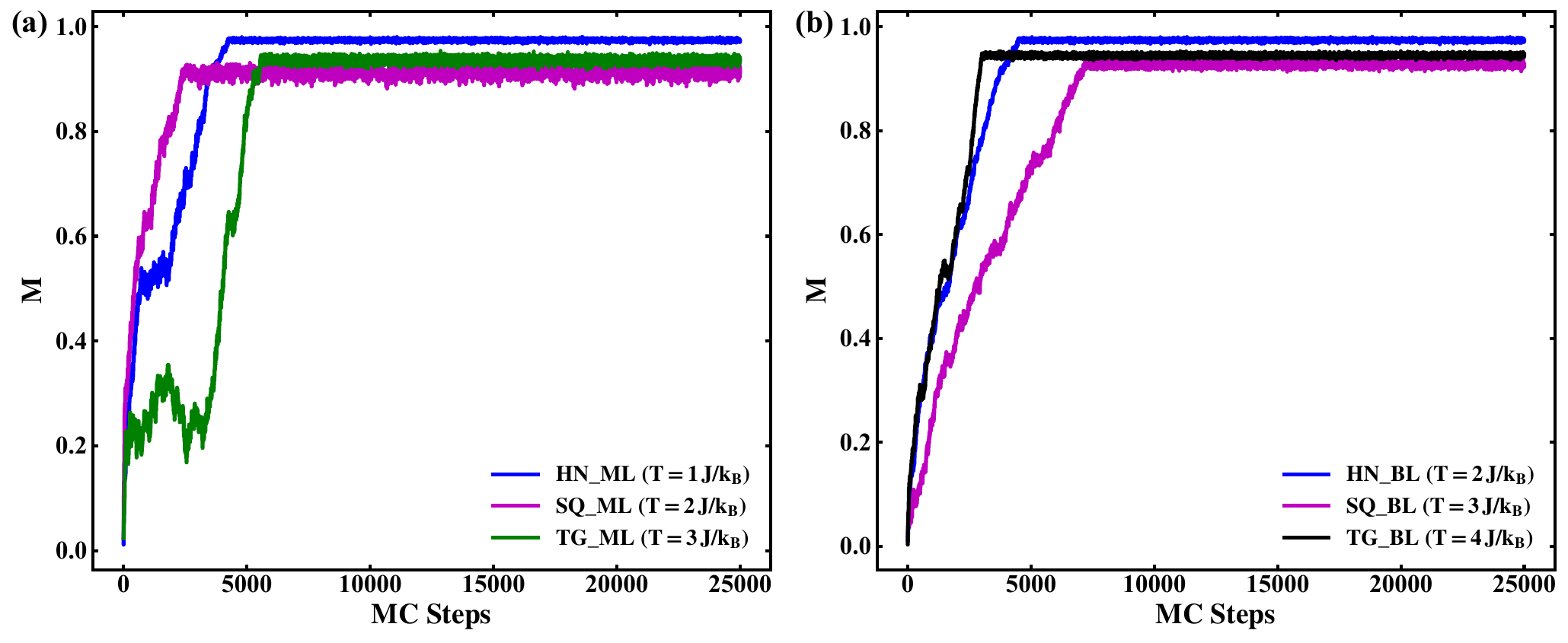}
		\caption{ Variation of $M$ with Monte Carlo (MC) steps plots are displayed for the square, honeycomb, and triangular lattices of (a) monolayer and (b) bilayer cases having $\mathscr{L}=128$ at fixed temperatures and $h=0$.}
		\label{Mg_SqHnHx}
	\end{figure}

\begin{figure}[H]
		\centering
		\includegraphics[width=16.0cm,height=5cm]{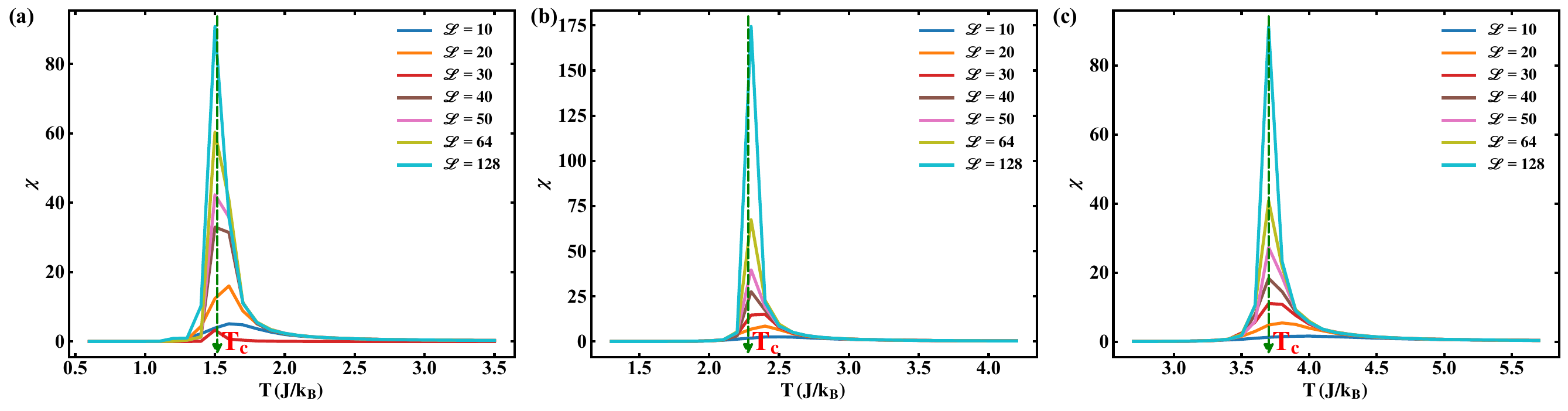}\\
		\caption{ $\chi$ vs. T plots for the (a) honeycomb, (b) square, and (c) triangular lattices for lattice sizes up to $\mathscr{L}=128$ while $h/J=0$. The transition temperature $T_c$ is indicated in each case, which is precisely determined by the Binder cumulant as shown in Fig.\ref{U_L_monolayer}.}
		\label{X_N_ml}
	\end{figure}

	We begin our numerical investigation by performing Monte Carlo simulations based on the Metropolis algorithm of the Hamiltonian Eq.\ref{Ham_Mono} and Eq. \ref{Ham_Bi} for the monolayer and bilayer lattices, respectively. We first consider $h=0$ and evaluate the magnetization (M) versus Monte Carlo steps using Eq.~\ref{eq_2} and note that for around $5\times10^3$ Monte Carlo steps the lattices equilibrate, as shown in FIG. \ref{Mg_SqHnHx}. We perform 1/3 of the total Monte Carlo steps for equilibration of the lattices. After that, to determine the thermodynamic quantities, typically additional $10^6$ Monte Carlo steps are considered for evaluating the thermodynamic averages. We next evaluate susceptibility ($\chi(T,0)$) by using Eq.~\ref{eq_3}. In Fig. \ref{X_N_ml}, we show the variation of the susceptibility ($\chi$) with temperature for different lattice sizes. In FIG. \ref{X_N_ml}, we show the variation of the susceptibility with T of different lattice sizes and find that the susceptibility exhibits a finite-size rounded peak near $T_c$, and the peak becomes prominent as the system size increases, indicating a second-order transition. We note that at a true second-order transition $\chi$ diverges to infinity for an infinite system, \textit{i. e.} $\chi(T,0) \sim |T-T_c|^{-\gamma}$ where $\gamma >0$ is the susceptibility exponent. For a finite system of size $L$, the divergence is cut off by finite size, and the correlation length $\xi(T)$ cannot exceed L. The correlation length $\xi(T) \sim |T-T_c|^{-\nu} $ with $\nu$ as the correlation exponent. For a finite system size of L near $T_c$, $\xi(T) \sim L$ and therefore the susceptibility ($\chi$) saturates at a finite peak value $\chi^{}_{max}(L) \sim L^{\nu/\gamma}$. We also note that there is no true divergence for a first-order phase transition, and the susceptibility remains finite in the thermodynamic limit (infinite system) at the transition temperature. In the first-order transition, the susceptibility exhibits a finite-size peak, but it scales as $\chi^{}_{max}(L)\sim L^d$ (volume scaling).
	
	For finite system sizes, using our numerical data, one can obtain the second-order critical temperature $T_c$ and the critical exponents $\gamma$ and $\nu$. For finite systems, the peak temperature $T_{\rm peak} (L)$ is not precisely at $T_c$, but it approaches $T_c$ as the system size increases, and the scaling relation is $T_{\rm peak}(L)-T_c \sim L^{-1/\nu}$. The intercept of the linear plot of $T_{\rm peak}(L)$ and $L^{-1/\nu}$ would provide the $T_c$. For 2D Ising model $\nu=1$ and $\gamma=7/4$. 
	
	We use the Jackknife method to calculate the error bars. The error bars are so small that they are dropped for the sake of the clarity of the images. Although for the sake of completeness, we show the thermodynamic quantities at different lattice sizes of $\mathscr{L}=$ 10, 16, 20, 24, 32, 64, and 128 with the error bars for the bilayer triangular lattice in the discussion ahead \ref{TQ_Sq_Hn_Hx_dl_mcs}.
	
	We perform the simulations for the above mentioned lattice sizes and we found that at lower field values the all quantities in general and especially the quantities important for the magnetocaloric effect like $(\partial M/ \partial T)_h$ and $C_M$ almost collapse above the field value of $h/J=0.08$ and the lattice size $\mathscr{L}\ge 16$, as shown in SI FIG. \ref{FSS_VLS}. As a result, the magnetocaloric quantities like $\Delta S_M$ and $\Delta T_{ad}$ do not show much variation with the increase in the lattice size, as shown in the supporting information (SI) FIG. \ref{FSS_VLS}. For the same reason, here, we show the results for the magnetocaloric quantities above the field ($h/J$) value of 0.08.
	
	\subsection{Binder cumulant and determination of $T_c$}
	\label{TC}
	
	\begin{figure}[H]
		\centering
		\includegraphics[width=15cm,height=10cm]{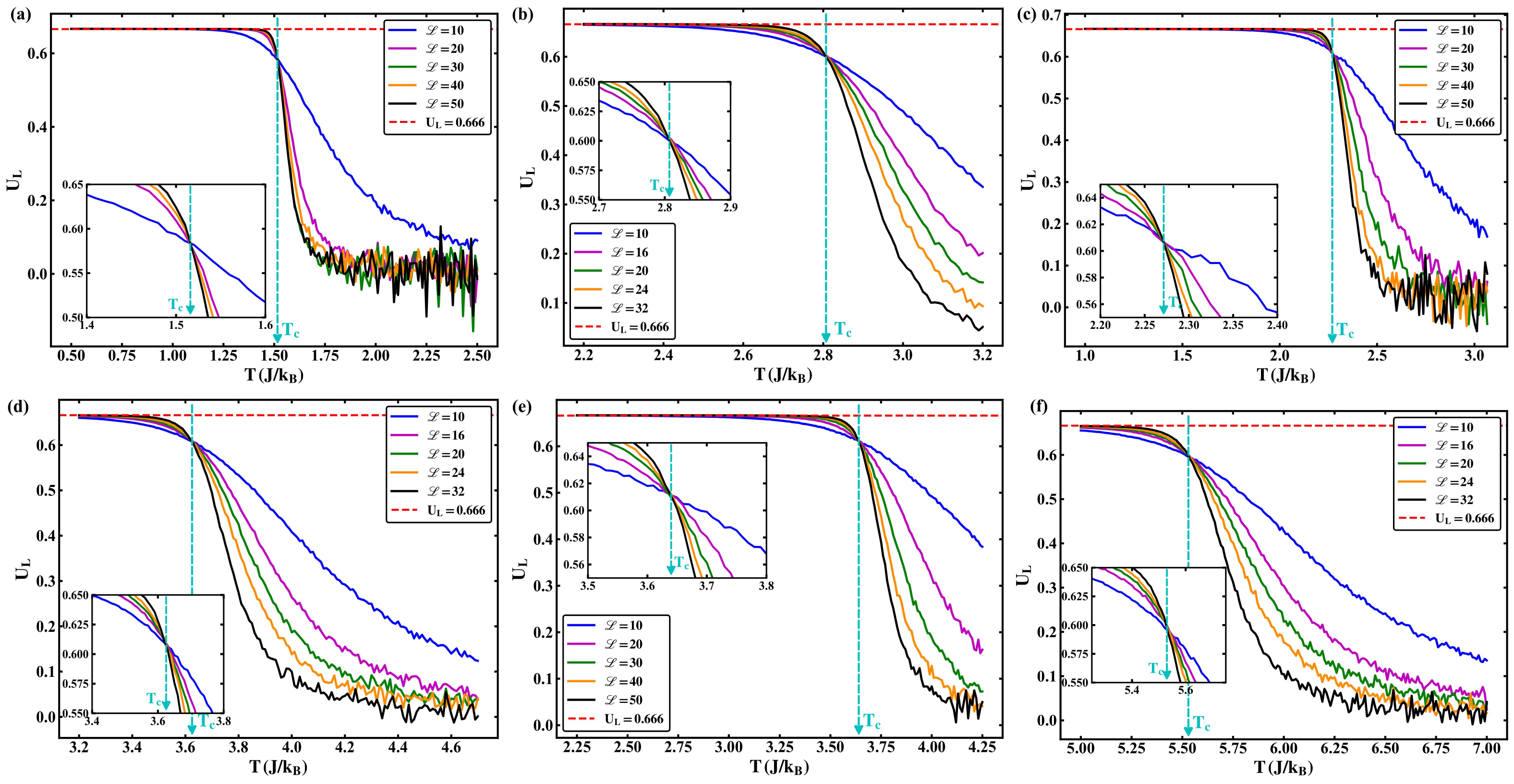}
		\caption{ $U_L$ vs. T plots of (a) honeycomb monolayer, (b) honeycomb bilayer, (c) square monolayer, (d) square bilayer, (e) triangular monolayer, and (f) triangular bilayer lattices for different $\mathscr{L}$ along the three axes. The cyan-colored vertical dashed lines indicate their transition temperatures ($T_c$), and the values of the transition temperatures are labeled on the plots.} 
		\label{U_L_monolayer}
	\end{figure}
	
	\begin{table}[H]
		\centering
		\caption{Comparison of transition temperatures ($T_c$) for monolayer and bilayer structures.}
		\begin{tabular}{|l|c|c|}
			\hline
			\textbf{Structure} & \textbf{Monolayer $T_c (J/k_B)$} & \textbf{Bilayer $T_c (J/k_B)$} \\ \hline
			Square & 2.27 & 3.62\\ \hline
			Honeycomb & 1.516 & 2.807 \\ \hline
			Triangular & 3.64 & 5.53 \\ \hline
		\end{tabular}
		\label{Table:Tab_1}
	\end{table}
	
	From the study of the variation of the Binder cumulant ($U_L$) with temperature, as shown in FIG. \ref{U_L_monolayer}, we present the transition temperatures of the three structures for the monolayer and bilayer cases. The intersection points of $U_L$ versus T curves for different system sizes of a particular lattice give the transition points of those systems. For the monolayer cases, we present the study of the lattice for different lattice sizes $\mathscr{L}=$ 10, 20, 30, 40, and 50. For the bilayer cases, we present the variation of $U_L$ with T studied for $\mathscr{L} =$ 10, 16, 20, 24, and 32. We use a Monte Carlo simulation based on the Wolff algorithm, with $3.33 \times 10^5$ Monte Carlo steps for equilibration and $10^6$ for calculating the magnetization and Binder cumulant. In our earlier studies, we also observed the Wolff algorithm to be a more effective way to determine the precise transition temperature for anisotropic Ising models \cite{iqbal-2024}. As discussed earlier, it easily bypasses critical slowing down and captures the transition efficiently. The Binder cumulant gives precise transition temperatures. The transition points for the three lattices are given in Table \ref{Table:Tab_1}. Among the three structures, the triangular lattice shows the highest transition temperature in both the monolayer and bilayer cases, followed by the square and honeycomb lattices. The transition temperatures increase as we go from their monolayer to their bilayer cases. The increase in the transition temperatures can be attributed to the increase in coordination numbers, which in turn increases the number of magnetic exchange interactions, resulting in stronger magnetic interactions between the spins. So, stronger thermal fluctuations are necessary to break the ferromagnetic order and for the phase transition.
	
	Finite-size effects are addressed using the Binder cumulant analysis. While thermodynamic quantities are shown for $\mathscr{L}=128$ for different lattices. The cumulant curves for different system sizes intersect at a nearly common temperature, providing a reliable estimate of the critical point in the thermodynamic limit. The stability of the crossing of the $U_L$ versus $ T$ curves with increasing lattice size indicates that finite-size effects are well controlled and do not significantly influence the reported thermodynamic behavior. At $T_c$, $U_L$ exhibits the universal value of $U_L$ denoted as $U^*_L \approx 0.61...$, which depends on the aspect ratio of the lattice \cite{malakis-2014,kamieniarz-1993}. We also observe this universality in $U_L$ exactly for the monolayer and bilayer square lattice, with a minor deviation on the honeycomb and triangular lattices, which can be attributed to the varying aspect ratio arising due to different $N_x$ and $N_y$ points as discussed above.
	
	In a separate study, we investigated the critical properties of the same lattices using Curie-Weiss mean-field theory (MFT) and single-spin cluster mean-field theory (SCMFT)\cite{iqbal-2026-mft}. The Curie-Weiss mean-field theory provides a qualitative description of phase transitions, but since it does not take into account thermal fluctuations, it leads to an overestimation of transition temperatures. The single-spin cluster mean-field theory, as compared to the Curie-Weiss mean-field theory, does a better job in approximating the transition temperatures for the three lattices. One can say that $T_c^{MC}<T_c^{SCMFT}<T_c^{MFT}$. Our $T_c$ of the monolayer cases are in close agreement with the $T_c$ values of Yamamoto \textit{et al.} \cite{yamamoto-2009}. 
	
	\subsection{Comparative analysis of thermodynamic properties of different lattices at zero field($h/J=0$).}
	
	\label{results}
	In the calculations of different thermodynamic quantities, first of all, 3.33$\times10^5$ steps are used for the system equilibration and discarded; after that, additional $10^6$ Monte Carlo steps are used while calculating the average values of the thermodynamic quantities. We performed all the simulations on various lattice sizes: $\mathscr{L}=$ 10, 20, 40, 50, 64, and 128 for the monolayer lattices; and $\mathscr{L}=$ 10, 16, 20, 24, 32, 64, and 128 for the bilayer lattices. Finally, the results shown here are for the lattice size $\mathscr{L}=128$ for both the monolayer systems and bilayer systems in all structures of square, honeycomb, and triangular lattices. We use the Jackknife method to calculate the errors, which are quite small, so they are dropped for better clarity of the graphs. The magnetization is measured in units of $\frac{J}{g\mu_B\mu_0}$, energy in units of $J$, specific heat in units of $k_B$, and the units of $\chi$ can be determined from Eq. \ref{eq_3}.
	
	\begin{figure}[t]
		\centering
		\includegraphics[width=16cm,height=7cm]{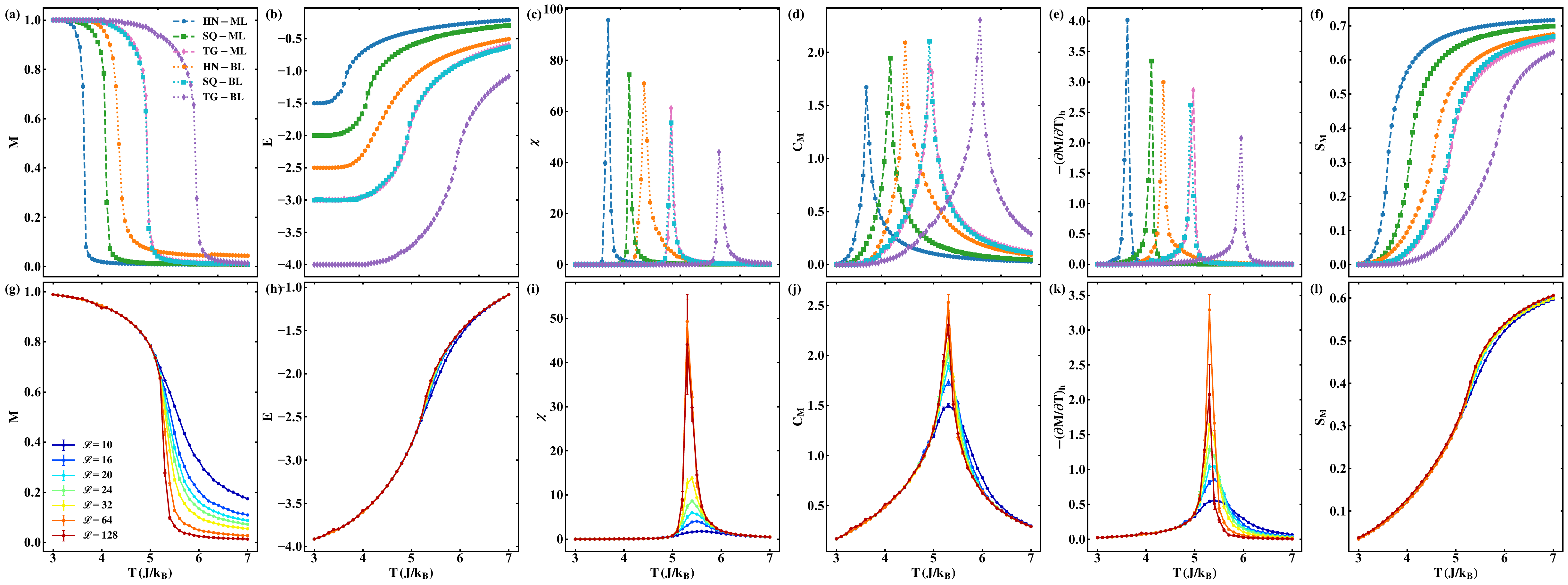}	
		\caption{(a) to (f): Thermodynamic and magnetic quantities for the monolayer and bilayer cases of the square, honeycomb, and triangular lattice for h/J = 0 and $\mathscr{L}=128$. (g) to (i): Thermodynamic quantities at different lattice sizes $\mathscr{L}$ with error bars calculated using the Jackknife method.}
		\label{TQ_Sq_Hn_Hx_dl_mcs}
	\end{figure}
	
	As we move from the honeycomb ($r=3$) to square ($r=4$) and triangular ($r=6$) lattice, the number of in-plane magnetic interactions for each spin increases. This effectively strengthens the magnetic interaction strength, which will enhance the collective magnetic ordering; hence, stronger thermal fluctuations are necessary to destroy the order, resulting in higher transition temperatures. With the addition of another layer, additional interlayer magnetic exchange interactions are introduced, which further enhance the magnetic strength of the lattice, further stabilizing the magnetic order; therefore, increases the transition temperature. FIG. \ref{U_L_monolayer} shows the precise transition temperatures for the six lattices as obtained from the Binder cumulant analysis, and the corresponding values are listed in Table \ref{Table:Tab_1}.
	
	First of all, we present the zero-field($h/J=0$) behavior of the thermodynamic quantities of the six lattices, as given in FIG. \ref{TQ_Sq_Hn_Hx_dl_mcs}. As shown in \ref{TQ_Sq_Hn_Hx_dl_mcs}(a), the variation of M with T depicts the phase transition from the ferromagnetic to paramagnetic state as the M value goes from $M\approx1$ to $M \approx 0$ at their corresponding $T_c$. We also observe that $T_c$ increases with the enhancement of magnetic interaction strength, which happens due to an increase in the intralayer and interlayer magnetic nearest-neighbor interactions. In the thermodynamic limit, $\chi_M$ diverges at $T_c$; for our finite-size lattices, $\chi_M$ shows a peak at $T_c$, shown in FIG. \ref{TQ_Sq_Hn_Hx_dl_mcs}(c), and we observe that its height decreases with increasing $r$ because stronger average spin interactions suppress relative critical fluctuations and stabilize the ordered phase. The quantity $(-\partial M/\partial T)_h$ is crucial in the study of magnetocaloric effects, and it also exhibits a peak at $T_c$. The $(-\partial M/\partial T)_h$, as shown in FIG. \ref{TQ_Sq_Hn_Hx_dl_mcs}(e) shows a peak at the transition temperature, and the peak values decrease as we go from the lower to higher number of nearest-neighbor magnetic interactions, as stronger magnetic interactions make the system more resistant to thermal disordering. Mean-field theories also certify the same behavior \cite{iqbal-2026-mft}. 
	
	FIG. \ref{TQ_Sq_Hn_Hx_dl_mcs}(b) presents the temperature dependence of the magnetic energy ($E$). The magnitude of $E$ increases as we move from the honeycomb to square and triangular lattices. Moreover, larger energy values are observed for the bilayer systems compared to the monolayer systems due to the additional interlayer interactions. In the ferromagnetic ground state, the energy per spin approaches $-rJ/2$(=$-r/2$, as $J=1$ in our case). As the system heats up, at corresponding $T_c$'s, the value of $E$ increases and approaches 0 at $T>>T_c$. The specific heat ($C_M$), as shown in FIG. \ref{TQ_Sq_Hn_Hx_dl_mcs}(d), shows a peak at the transition temperature, indicating a large energy fluctuation, indicating a breakdown in the magnetic ordering at $T_c$. Since, with the increase in the coordination number, the magnetic exchange interaction is stronger, which makes thermal excitation of the spins harder; hence, the $C_M(T,h)$ value decreases, but the variations observed are quite marginal. The magnetic entropy per spin ($S_M$), as shown in FIG. \ref{TQ_Sq_Hn_Hx_dl_mcs}(f), is zero at low temperatures due to the highly ordered ferromagnetic state and gradually increases toward the limiting value $k_Bln(2\sigma+1) \approx$ 0.693$k_B$ (for $\sigma=1/2$) after the phase transition. The variation of $S_M$ with T shows shallower slopes for higher coordination number, which can be ascribed to increasing magnetic interaction strengths increasing the competition for thermal fluctuations.
	
	\subsubsection{Comparative analysis of the magnetocaloric properties of different lattices.}
	\label{MCE_SU} 
	
	Following the behavior predicted by the relation given by Eq.\ref{eq_8} as well as the zero-field $(-\partial M/\partial T)_h$ results as discussed above and shown in FIG. \ref{TQ_Sq_Hn_Hx_dl_mcs}(e), one can anticipate larger $-\Delta S_M$ for the honeycomb as compared to the square and triangular lattices, because larger values of $(-\partial M/\partial T)_h$ shall produce larger $-\Delta S_M$. Similarly, one can expect $-\Delta S_M$ values for the monolayer systems to be larger than the bilayer systems. Further, the M versus T plots also indicate that the transition from the ferromagnetic to paramagnetic region is steeper for (1) lattices with a lower coordination number than higher ones, (2) monolayer lattices than bilayer lattices. This can be attributed to an increase in magnetic interaction strength and the resulting broadening in the competition region of the thermal fluctuations and magnetic ordering. The lattices with lower coordination numbers have lower magnetic interaction strength, so the thermal fluctuations succeed over magnetic interactions in a smaller temperature range, but in the lattices where the magnetic strength is stronger, the competition extends over a longer temperature range.
	
\begin{figure}[H]
		\centering
		\includegraphics[width=16cm,height=6cm]{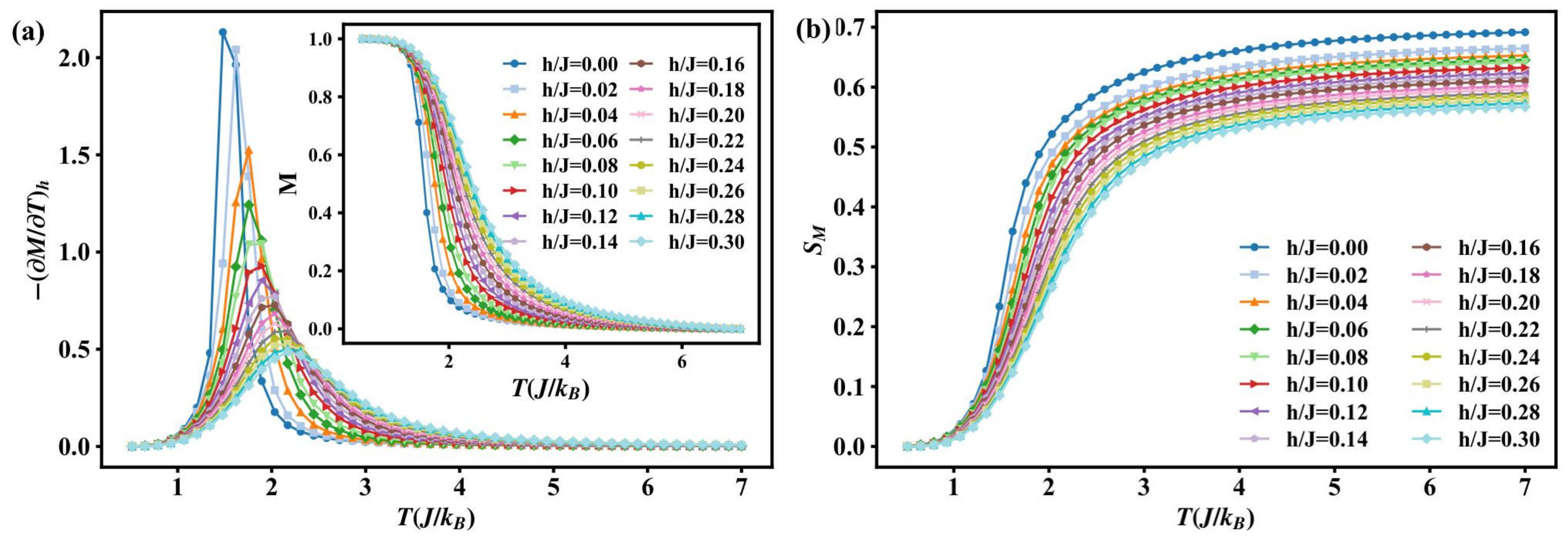}
		\caption{The variation of the (a) $-\frac{\partial {M}}{\partial {T}}$ and $M$(in the inset) and (b) $S_M$ at different fields for the monolayer honeycomb lattice from the Monte Carlo simulation study. We found similar field and temperature dependence in the square and triangular (monolayer and bilayer) lattices at their respective $T_c$'s.}
		\label{Mh_MCS}
	\end{figure}

	As shown in FIG. \ref{Mh_MCS}(a), for the honeycomb lattice, the variation of M with T for different fields shows that as the field increases, the curves become progressively smoother and the phase transition from the ferromagnetic region to the paramagnetic state extends to a longer temperature range. Similarly, the $-(\partial M/ \partial T)_h$ versus T curves broaden, and their peak values decrease with the increase in the field. This occurs because the external magnetic field stabilizes spin alignment and suppresses thermal disorder, thereby smoothing the magnetic transition. The monolayer square and triangular lattices also exhibit similar behavior. As discussed earlier in the zero-field ($h/J=0$) case, their $-(\partial M/ \partial T)_h$ peaks are lower compared to the honeycomb lattice (due to the addition of the intra-layer nearest-neighbor interaction), and the peak values decrease further as the field increases. As we go from monolayer to bilayer, the additional interlayer nearest-neighbor interaction further enhances the magnetic ordering tendency, resulting in an additional reduction in the peak heights of $-(\partial M/ \partial T)_h$, which decreases further with the increase in field. $\Delta S_M$ is calculated from the $-(\partial M/ \partial T)_h$ using high-accuracy numerical integration schemes. 
	
	Figure \ref{Mh_MCS}(b) illustrates the temperature dependence of the $S_M$ at different fields. At $h/J = 0$, the entropy rises sharply from zero and reaches its limiting value $\approx 0.693 k_B$, corresponding to the maximum disorder for a spin-$1/2$ system. As the field increases, the phase transition is progressively smeared, and the sharp increase in entropy evolves into a smoother variation over a broader temperature range. The field-induced reduction in entropy from zero to non-zero field defines $\Delta S_M(T,h) = S_M(T,h) - S_M(T,0)$. In the paramagnetic state, since the field enforces spin alignment, the entropy curves shift downward with increasing field strength. At sufficiently high temperatures($T>>T_c$), however, thermal fluctuations dominate, and $S_M$ reaches its limiting value of $\approx 0.693 k_B$, independent of the applied field.
	
\begin{figure}[H]
		\centering
		\includegraphics[width=16cm,height=18cm]{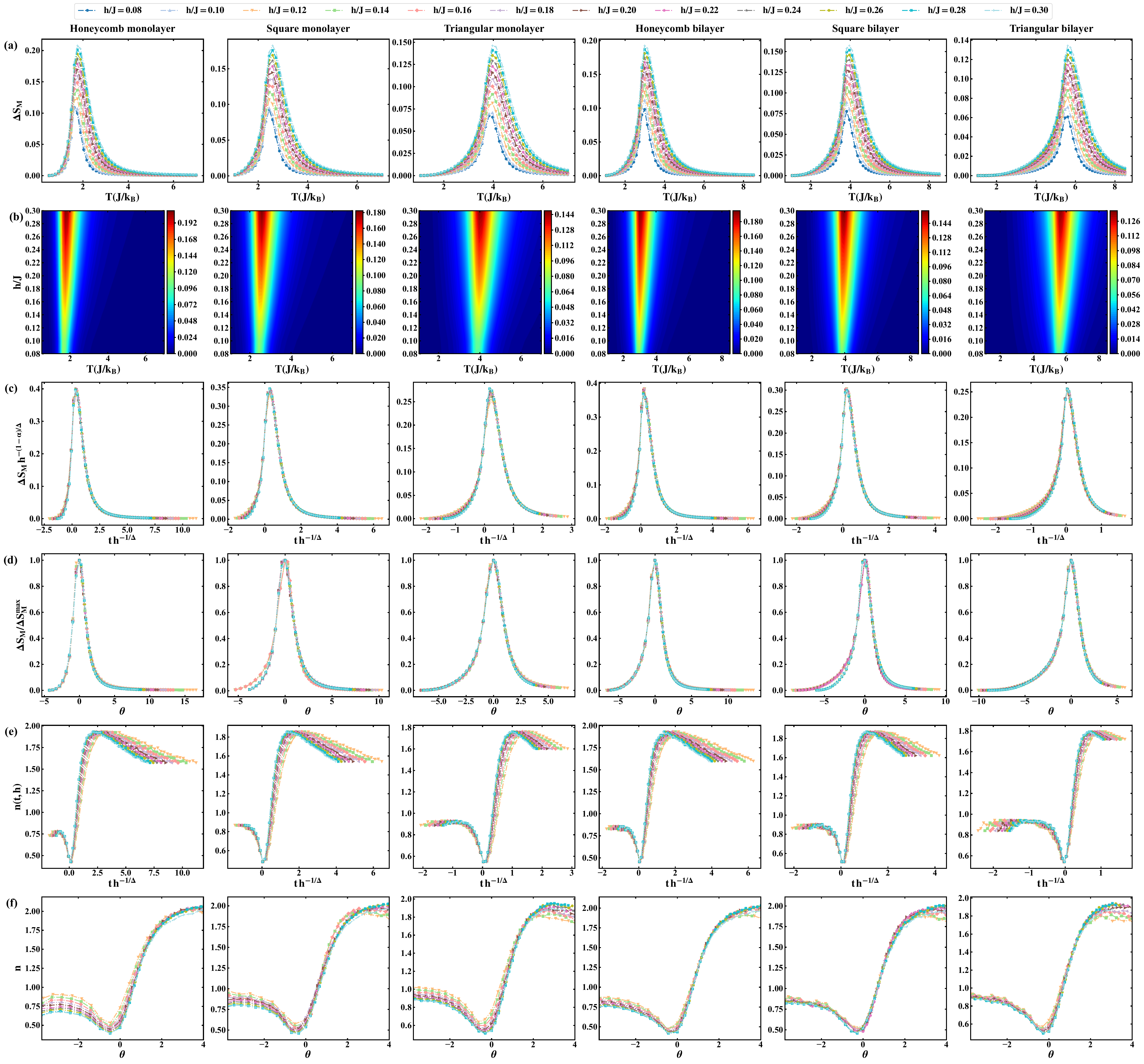}
		\caption{Temperature and field dependence of $-\Delta \mathrm{S}_{\mathrm{M}}$ as (a) line plots and (b) color plots; variation of (d) $\Delta S_M/\Delta S_M^{Max}$ and the (f) exponent $n$ with $\theta$; and scaling of the (c) $\Delta S_M$ and the (e) exponent $n$ as per their scaling laws at different fields; for the monolayer and bilayer honeycomb, square and triangular lattices.}
		\label{dels_rcp_MCS}
	\end{figure}

	FIG. \ref{dels_rcp_MCS}(a) and (b) show the variation of $\Delta S_M$ with T at different magnetic fields for monolayer and bilayer cases, in the form of line plots and color-maps, respectively. In the calculation of $\Delta S_M$, we use the quantity $-(\partial M/ \partial T)_h$ as given by Eq. \ref{eq_8}. This way, from the behavior of $-(\partial M/ \partial T)_h$, one can anticipate peak-like behavior in $\Delta S_M(T, h)$ as well. Like the variation of $-(\partial M/ \partial T)_h$ with T, the $-\Delta S_M(T, h)$ exhibits a peak near $T_c$, indicating a phase transition. Below $T_c$, the system remains ferromagnetically ordered, in which the spins are largely aligned in a single direction, resulting in low $S_M$. Above $T_c$, as the thermal fluctuations progressively overcome the magnetic interaction, the system transitions to a paramagnetic state in which the spins are thermally disordered, leading to high $S_M$. At $T_c$, the system shows a rapid transition from a highly ordered and low-entropy state to a disordered and high-entropy state, leading to a pronounced peak in $\Delta S_M$ near $T_c$ representing this rapid change in magnetic order. 
	
	As discussed earlier about the zero-field results in FIG. \ref{TQ_Sq_Hn_Hx_dl_mcs}(e), the magnitude of the peak value of $-(\partial M/ \partial T)_h$ decreases as the coordination number increases. We observe similar behavior in the variation of $-\Delta S_M$ with different field and T studies. With the increase in the field, the $-\Delta S_M$ values increase because stronger external fields induce greater spin alignment and consequently larger entropy reduction. As we go from the honeycomb to the square and triangular lattice, the coordination number increases, which enhances magnetic interaction strength and suppresses thermal fluctuations, resulting in a decrease in the $-\Delta S_M$ peak values. For the same reason, at a given field, the bilayer $-\Delta S_M$ peak is smaller than that of the corresponding monolayer. Similar behavior has also been reported within mean-field theory studies \cite{iqbal-2026-mft}.
	
\begin{figure}[H]
		\centering
		\includegraphics[width=16cm,height=12cm]{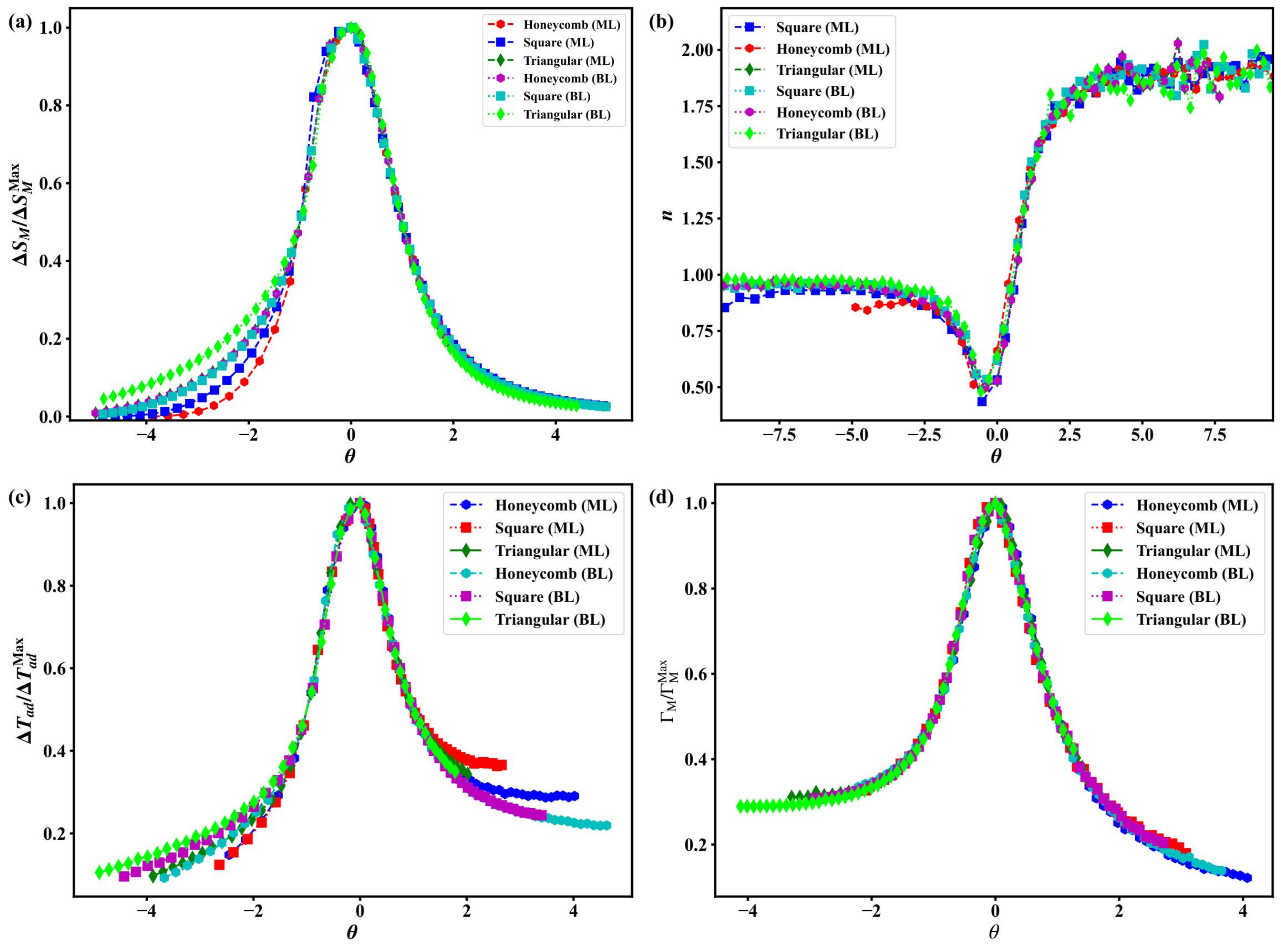}
		\caption{Variation of (a) $\Delta S_M/\Delta S_M^{Max}$, (b) exponent $n$, (c) $\Delta T_{ad}/\Delta T_{ad}^{Max}$ and (d) $\Gamma_{M}/\Gamma_{M}^{Max}$ with $\theta$ plots of the square, honeycomb, and triangular lattices for their monolayer and bilayer cases at $h/J$=0.15 (except exponent $n$ which is at $h/J$=0.08) indicating the universal behavior and scaling in the magnetocaloric properties across these lattices.}
		\label{scaling_n_ml_bl}
	\end{figure}

	As discussed above, Franco \textit{et al.} proposed universal scaling behavior of $\Delta S_M$. According to this approach, $\Delta S_M(T, h)$ measured at different magnetic fields can be rescaled to a single universal master curve independent of the field. FIG. \ref{dels_rcp_MCS}(d) shows the scaling behavior for the lattices of the normalized magnetic entropy change, \textit{i. e.} $\Delta S_M / \Delta S_M^{Max}$ as a function of the reduced temperature parameter $\theta$, where $\theta$ is defined by Eq. \ref{eq_theta}. The reference temperatures $T_{r1}$ and $T_{r2}$ correspond to $\theta=-1$ and $\theta =+1$ after normalization. $\Delta S_M^{Max}$ is the maximum value of $\Delta S_M$ that is near the transition temperature. By definition $\Delta S_M / \Delta S_M^{Max} = 1$ at $T_c (\theta =0)$ and it decreases as $\theta$ increases or decreases. The results of $\Delta S_M / \Delta S_M^{Max}$ versus $\theta$ corresponding to different field values collapse onto a single master curve for each lattice individually. Such scaling behavior is characteristic of second-order magnetic phase transitions and indicates that the systems exhibit the same underlying critical scaling behavior despite the variation in magnetic field and are driven by the same phenomenological behavior\cite{franco-2006,franco-2008}. Furthermore, as shown in FIG \ref{scaling_n_ml_bl}(a), a similar collapse is also observed for $\Delta S_M / \Delta S_M^{Max}$ curves of all six lattices, at a particular field. This suggests that the universal scaling and phenomenological behavior of $\Delta S_M$ extends not only across different magnetic fields but also across different lattice geometries. Although their $T_c$'s vary due to differences in coordination number and interaction strength, all of the six lattices appear to follow the same scaling characteristics associated with the two-dimensional Ising universality class.
	
	\begin{figure}[H]
		\centering
		\includegraphics[width=16cm,height=8cm]{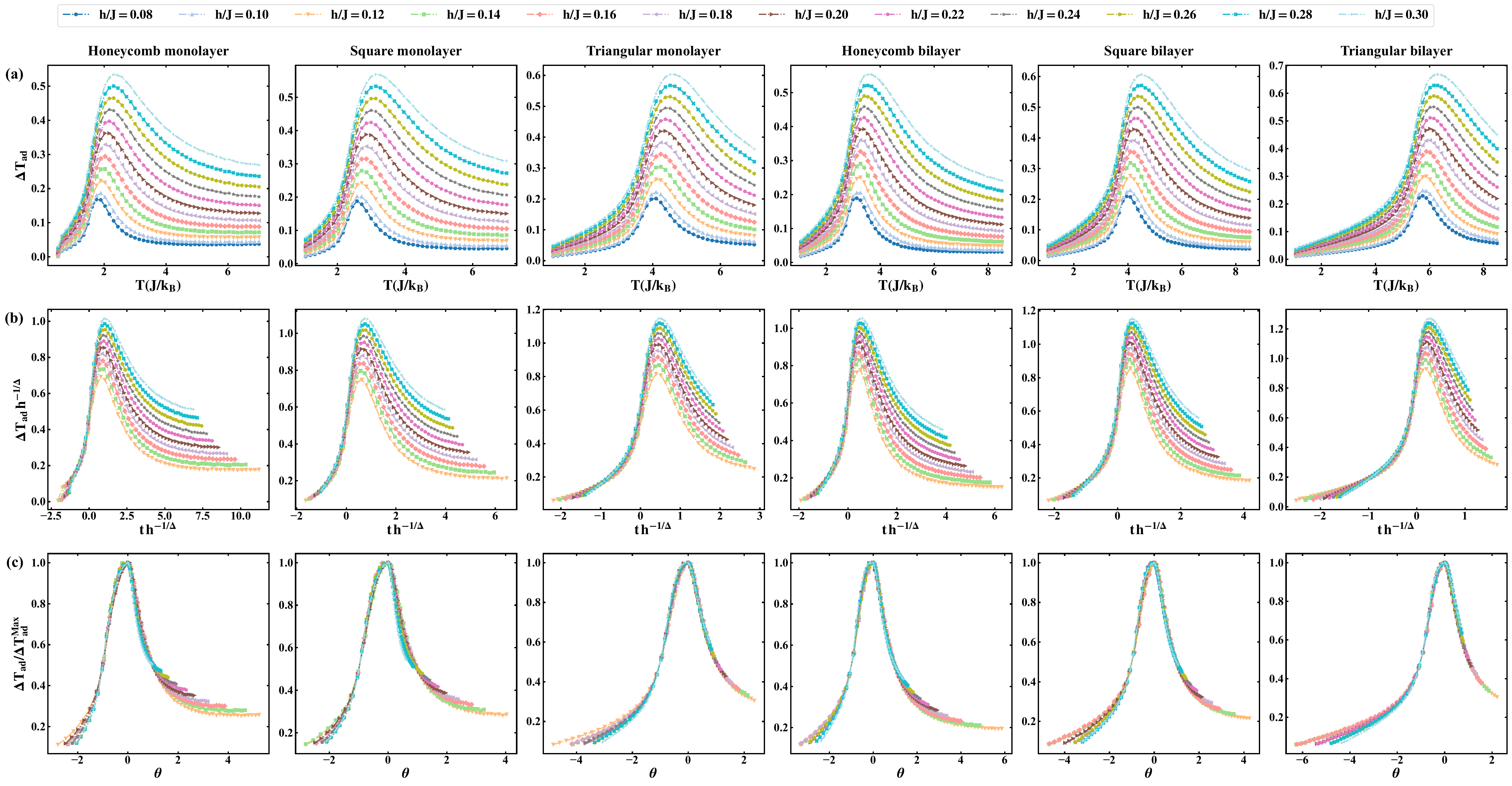}
		\caption{(a) Temperature and field dependence of $\Delta \mathrm{T}_{ad}$; scaling of (c)$\Delta T_{ad}$ as per the scaling law; and (c) variation of $\Delta T_{ad}/\Delta T_{ad}^{Max}$ with $\theta$; and at different fields; for the monolayer and bilayer honeycomb, square and triangular lattices.}
		\label{delT_all_MCS}
	\end{figure}
	
	FIG. \ref{dels_rcp_MCS}(e) shows the scaling behavior of the field exponent $n$, as defined in Eq. \ref{n_T}, for the monolayer and bilayer honeycomb, square, and triangular lattices. At temperatures below $T_c$, the critical exponent $n$ approaches 1, indicating an approximately linear field dependence of the magnetic entropy change in the ferromagnetic region. Moreover, at temperatures above $T_c$, the exponent $n$ approaches 2, reflecting the quadratic field dependence of the entropy change predicted by the Curie–Weiss law in the paramagnetic regime \cite{spaldin-2010} where $S_M \propto h^2$. At the critical temperature, Oesterreicher and Parker predicted the mean-field value $n(T_c) = 2/3$ for the Heisenberg model \cite{oesterreicher-1984}. However, the exact value is directly related to the critical exponents, as expressed in Eq. \ref{n_T_c}, and can be deduced from the critical exponents of the material and model \cite{franco-2006}. For our case, using the critical exponents of the two-dimensional Ising model having $\beta= 0.125$ and $\gamma = 1.75$\cite{yin-2009}, at $T_c$, the exponent $n(T_c) \approx 0.533$ \cite{yin-2009}, which is consistent with the present simulation results. We observe that the scaling laws and universal behavior hold for all magnetic fields considered here and for all lattices individually. Interestingly, as shown in FIG. \ref{scaling_n_ml_bl}(b), the collapse of $n$ versus $\theta$ curves of the six lattices further supports the existence of universal scaling and phenomenological behavior across different lattice structures. This suggests that for the range of parameters studied, the interlayer coupling does not significantly alter the shape of the scaling function, or that all six lattices remain in the same effective universality class. The results are further supported by the reports of the mean-field studies of the same lattices \cite{iqbal-2026-mft}. The $n$ collapse further confirms that the scaling hypothesis holds because the temperature‑dependent exponent $n$ converges to the theoretically expected limits. However, above some particular $h/J$ value, depending on the $r$ value of the lattice, we observe some deviations as discussed ahead.
	
	\begin{figure}[H]
		\centering
		\includegraphics[width=16cm,height=9cm]{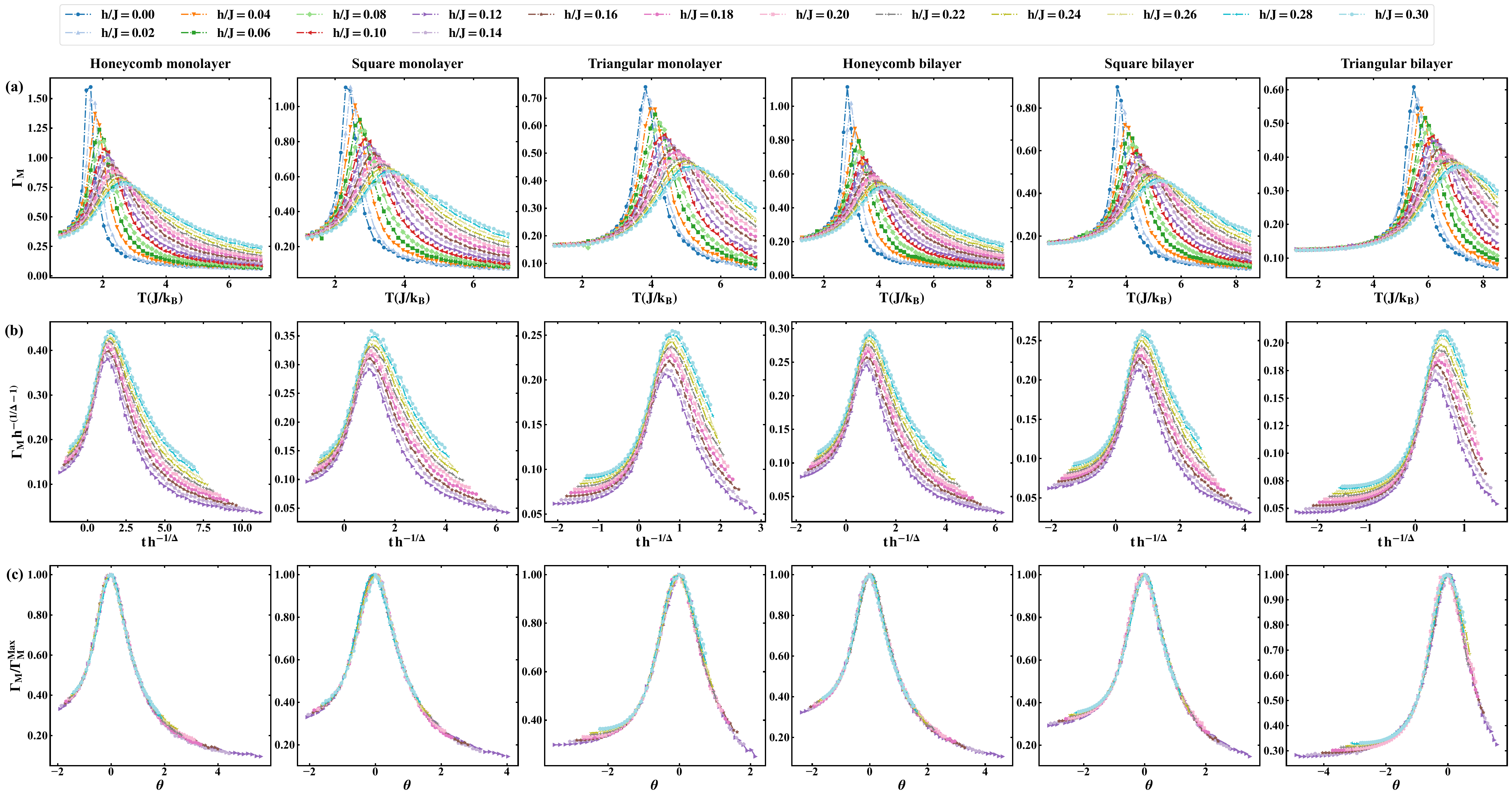}
		\caption{(a) Temperature and field dependence of $\Delta T_{ad}$; (b) scaling of $\Gamma_{M}$ as per the scaling laws; and (c) variation of $\Gamma_{M}/\Gamma_{M}^{Max}$ with $\theta$ at different fields; for the monolayer and bilayer honeycomb square and triangular lattices.}
		\label{Gamma_all_MCS}
	\end{figure}
	
	The color plots (FIG. \ref{Figure_1}(b)) provide a complementary visualization of $\Delta S_M$ behavior. Near the high-field region around the corresponding transition temperatures, a concentrated dark red/orange region is observed, representing the maximum values of $\Delta S_M$, consistent with the sharp peaks observed in the line plots. As the magnetic field decreases or we move away from $T_c$, the color gradually regresses from red to orange, yellow, and green, reflecting the gradual suppression and broadening of the critical anomaly--indicating a supercritical region.
	
	Now, coming to the discussion of $\Delta T_{ad}$, FIG. \ref{delT_all_MCS}(a) shows the variation of $\Delta T_{ad}$ with T, for all three lattices and for monolayer and bilayer cases. As in the magnetocaloric study, $\Delta T_{ad}$ is obtained from $\Delta S_{M}$; the behavior of the two quantities is quite similar. $\Delta T_{ad}$ shows a peak at the transition point, but contrary to $\Delta S_M$ behavior, the peak values of $\Delta T_{ad}$ increase with the increase in the coordination number as we move from honeycomb to square and triangular lattices, and as we go from the monolayer to the bilayer case. The increase in $\Delta T_{ad}$ with the increase in the coordination number can be understood as follows. As given in Eq. \ref{eq_10}, here we have an extra factor of $T/C_M(T,h)$ multiplied by $(\partial M/ \partial T)_h$ in the integral. The peak of $(\partial M/ \partial T)_h$ is near $T_c$. So, $\Delta T_{ad}$ peak is also near $T_c$. First of all, as discussed earlier, with the increase in coordination number, $T_c$ increases, which will increase the factor $T/C_M(T, h)$. Secondly, as far as $C_M(T, h)$ is concerned, it quantifies the amount of energy the material can absorb for a given temperature increase. With the increase in the coordination number, the number of nearest neighbor interactions increases, which makes the magnetic exchange interaction stronger. So, with the increase in the coordination number, thermal excitation of the spins becomes harder. Thus, $C_M(T,h)$ value decreases. Hence, the factor $T/C_M(T, h)$ increases with the increase in the coordination number, and that results in an increase in the peak values of $\Delta T_{ad}$ with the increase in the coordination number. In our studies, we observe that the decrement in $C_M$ is quite marginal; the increment in $T_c$ plays the main role in the increment in $\Delta T_{ad}$.
	
	FIG. \ref{Gamma_all_MCS} shows the magnetic Gr\"{u}neisen parameter ($\Gamma_{M}$), as a function of the temperature and applied magnetic field, presented respectively as a line plot with parametric curves for the six lattices. At the low field values, $\Gamma_{M}$ exhibits a sharp, narrow peak that reaches the peak value near the $T_c$ of the corresponding lattices. As field intensity increases, the peak shifts monotonically to higher temperatures, and its maximum height steadily decreases and broadens. Furthermore, as we move from the honeycomb to the square and triangular lattice or from monolayer to their corresponding bilayer lattices, the peak values decrease. The behavior is qualitatively similar to that of $\Delta S_M$ and can be understood using arguments consistent with the earlier analysis of the $-\Delta S_M$. With the increase in the coordination number, the magnetic interactions become stronger and $\Gamma_{M}$ decreases, as stronger magnetic coupling suppresses thermal spin fluctuations, thereby reducing the magnitude of $\Gamma_{M}$, which characterizes the magnetocaloric response of the system under adiabatic field variation. 
	
	As presented above, like $\Delta S_M/\Delta S_M^{\max}$, the normalized quantities $\Delta T_{\text{ad}}/\Delta T_{\text{ad}}^{\max}$ and $\Gamma_M/\Gamma_M^{\max}$ collapse onto single master curves when plotted against the reduced temperature $\theta$. This collapse occurs both for different fields within a given lattice and for all lattice structures at a fixed field. By the same reasoning applied to $\Delta S_M$, this scaling behavior provides further evidence for the second‑order nature of the phase transitions and the existence of the same universal scaling and phenomenological behavior across all the lattices. As shown in Fig \ref{scaling_n_ml_bl}(c) and (d), similar to $\Delta S_M$, $\Delta T_{ad}$ and $\Gamma_{M}$ with respect to $\theta$ at $h/J=0.15$ for the six systems exhibit a collapse, providing further indications that the same scaling characteristics are followed by the six lattices. Moreover, because $\Delta T_{\text{ad}}$ and $\Gamma_M$ are related to $\Delta S_M$ through the specific heat, whether these derived quantities would also obey scaling and exhibit universality requires explicit demonstration because the scaling does not hold for these quantities as per their scaling laws (as discussed ahead). The collapse across all geometries confirms that the three monolayer lattices belong to the same two‑dimensional Ising universality class, and the results remain consistent with the earlier findings for $\Delta S_M$.
	
	\begin{figure}[H]
		\centering
		\includegraphics[width=15cm,height=10cm]{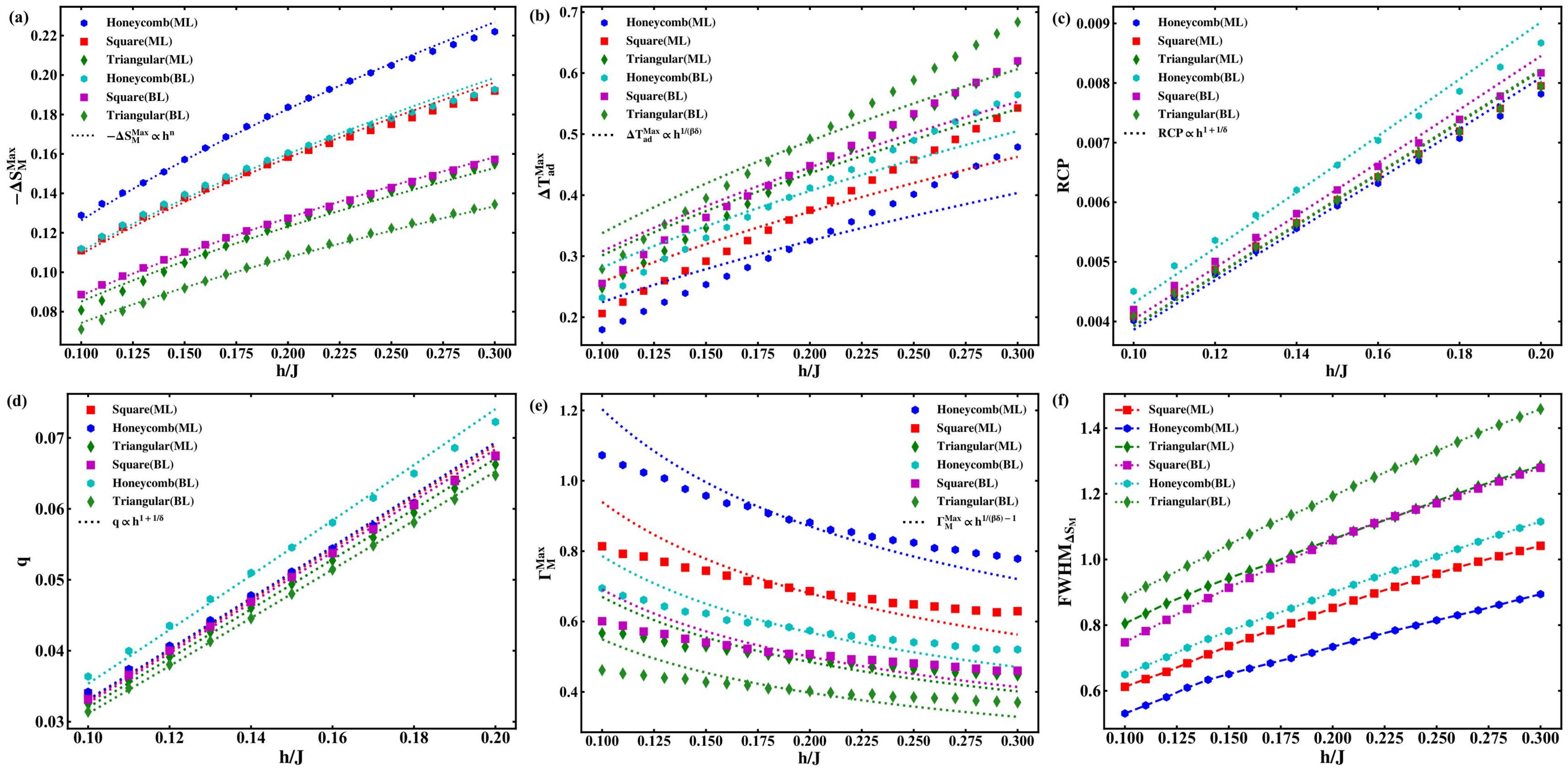}
		\caption{The field(h/J) dependence and the power laws of (a) $\Delta S_M^{Max}$, (b) $\Delta T_{ad}^{Max}$, (c) RCP, (d) $q$, (e) $\Gamma_{M}^{Max}$ and (f) $FWHM_{\Delta S_{M}}$ of $\Delta S_{M}$ curves for the monolayer(ML) and bilayer(BL) lattices of the honeycomb, square and triangular lattice structures. As seen earlier, $\Delta S_M^{Max}$ and $\Delta T_{ad}^{Max}$ are near $T_c$. The markers show the actual data for the different structures as discussed in Eqs. \ref{eq_8}-\ref{eq_10}, and the dotted lines show the fitting lines for the corresponding structures as per their power laws as discussed in Eqs. \ref{h^n}-\ref{deltaT_scaling}.}
		\label{power_laws}
	\end{figure}
	
	FIG. \ref{power_laws} shows the field dependence of the maximum of $\Delta S_M$, the maximum of $\Delta T_{ad}$, RCP, and $q$ for the studied structures. Here, we will discuss the variation of the actual values obtained numerically (shown by the markers), and then, in the following paragraph, we will discuss their power laws (shown by the dotted line). As shown in FIG. \ref{power_laws}(a) and as previously discussed in relation to FIG. \ref{dels_rcp_MCS}(a) at a particular field, the honeycomb lattice shows the largest values of $\Delta S_M$ among the three lattice structures because of the smaller number of nearest neighbors, and the values decrease as the number of nearest neighbors increases--when we move from the honeycomb to the square and triangular lattices or when we advance from monolayer to bilayer systems. This behavior originates from its lower coordination number and comparatively weaker magnetic coupling. In contrast, the behavior of $\Delta T_{ad}$, shown in Fig. \ref{power_laws}(b), is opposite to that of $\Delta S_M$ as observed previously. As shown in FIG. \ref{power_laws} (c) and (d), the quantities RCP and $q$ for the six structures exhibit only weak variation among the six lattice structures, yielding nearly identical values. This behavior can be understood from $\Delta S_M$ curves shown in FIG. \ref{dels_rcp_MCS}. As shown in Fig. \ref{power_laws}(f), although with the increase in the number of nearest neighbors, the peak values of $\Delta S_M$ decrease, but the $FWHM_{\Delta S_M}$ of the corresponding $\Delta S_M$ curves increases. As a result, the reduction in peak height is largely compensated by the broadening of the curves, leading to nearly unchanged values of RCP and $q$ across the different lattice geometries. Broader magnetocaloric peaks can be considered advantageous for practical refrigeration applications because they extend the operating temperature range.
	
	Fig. \ref{power_laws}(e) shows the extracted peak maxima of $\Gamma_M^{Max}$ as a function of the reduced magnetic field ($h/J$) across different 2D lattice geometries, and it has been observed that $\Gamma_M$ is suppressed with the increase in the magnetic field. As we go from the honeycomb to the square and triangular lattice, $\Gamma_M$ decreases. Similarly, as we go from monolayer to bilayer lattices, $\Gamma_M$ decreases. A lower coordination number means fewer nearest-neighbor bonds to stabilize magnetic order against thermal fluctuations. Consequently, the honeycomb lattice experiences much more volatile thermal fluctuations near its critical region compared to the more rigidly connected triangular lattice. Because $\Gamma_M$ is a direct probe of thermodynamic fluctuations, it naturally reaches significantly higher values in lattices with lower $r$. Adding a second layer introduces interlayer exchange coupling, effectively increasing the effective coordination number of each spin. This extra dimension of connectivity stabilizes the system, dampens critical thermal fluctuations, and reduces the $\Gamma_M^{Max}$.
	
	The scaling laws for the three thermodynamic quantities \textit{i. e.} $\Delta S_M$, $\Gamma_{M}$, $\Delta T_{ad}$, RCP and $q$ are discussed in section \ref{scaling_and_universality}. The power laws are followed by the structures in the low field region. $\Delta S_M$ scales according to the power law for a larger range of $h/J$ values. The scaling law for $\Delta T_{ad}$ gives wrong predictions of the scaling. This can be ascribed to the fact that $\Delta T_{ad}$ depends on the full entropy curves (including the non-singular part) of the specific heat and entropy \cite{smith-2014}. For the same reasons, the quantity $\Gamma_{M}$ also shows deviations from the scaling laws. Theoretically, RCP and $q$ scale as per the same scaling law, but we observe that $q$ follows the scaling law a little better than the RCP. In our investigations, we found that as the field increases, the magnetocaloric quantities start deviating from the predictions of the scaling law. This is due to the saturation of the magnetization as discussed in the following paragraphs. This deviation decreases as the number of nearest neighbors increases. The cause of the deviation is also discussed below. The results of mean-field theory studies lead to the same behavior \cite{iqbal-2026-mft}.
	
	\section{Hysteresis Loop}
	\label{Hyst}
	
	The hysteresis loops for the monolayer and bilayer cases of the honeycomb, square, and triangular lattices are plotted at different temperatures. The number of Monte Carlo steps is 5$\times 10^3$, and the size is $\mathscr{L}=20$ for both monolayer and bilayer the bilayer lattices. 
	
\begin{figure}[H]
		\centering
		\includegraphics[width=16cm,height=5cm]{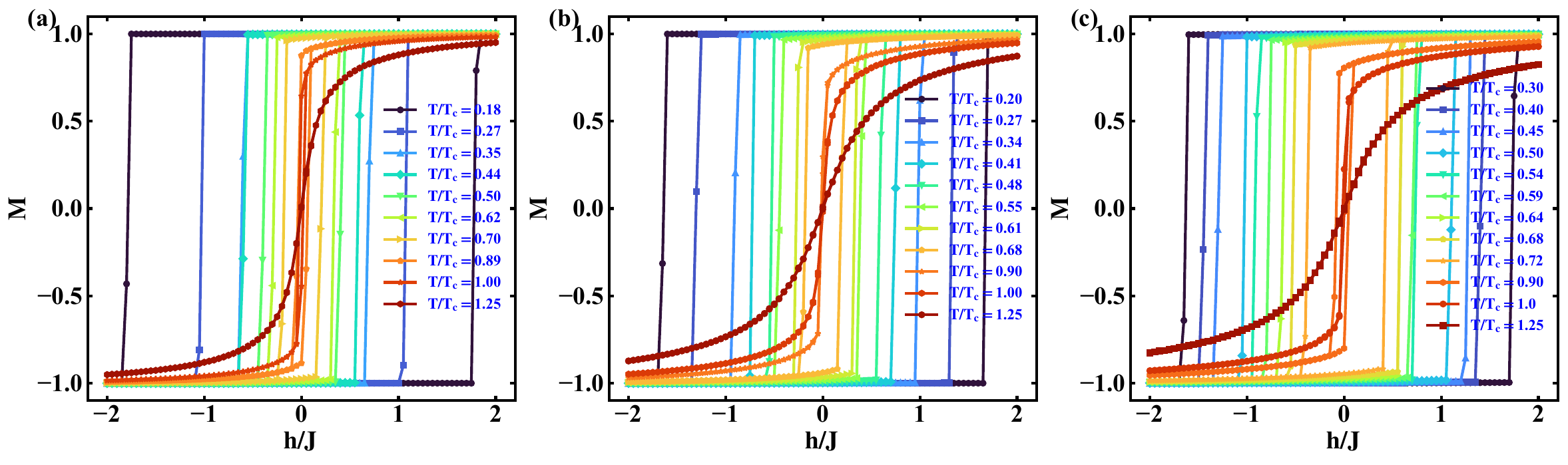}
		\caption{Hysteresis loop plots of the bilayer (a) honeycomb, (b) square, and (c) triangular lattices. We found that their monolayer counterparts also show similar behavior.}
		\label{Hyst_img}
	\end{figure}

	FIG. \ref{Hyst_img} showcases the temperature-dependent magnetic hysteresis loops ($M$ vs. $h/J$) of typical ferromagnetic systems at different temperatures in terms of their $T_c$. The evolution of these loops perfectly illustrates the distinction between a first-order and a second-order magnetic phase transition. At temperatures well below $T_c$, the magnetization ($M$) undergoes a sudden, abrupt jump from $-1$ to $+1$. This discontinuous jump signifies a first-order transition. The system possesses two stable energy minima (spontaneous magnetization with all magnetic moments pointing up or down). Applying the field, at $T<T_c$, forces a sharp, discontinuous switch between these states, creating a wide hysteresis loop where the path forward differs from the path backward. As the temperature increases, thermal fluctuations begin to destabilize the ordered ferromagnetic state. The hysteresis loop progressively narrows, and the coercive field (the field required to bring $M$ to zero) shrinks toward zero. At $T \to T_c$, the two sides of the hysteresis loop merge into a single, continuous, reversible curve passing through the origin. The magnetization changes smoothly without any discontinuous jumps, marking the classic continuous behavior of a second-order phase transition. This behavior continues for $T>T_c$
	
	The hysteresis loop shows the energy loss in the $M-h$ cycles. The magnetocaloric effect is an ideally reversible thermodynamic process that converts magnetic work into heat. In a first‑order phase transition, the transition is not hysteresis-free, as strong hysteresis appears at $T_c$ \cite{gutfleisch-2016,franco-2017}. But for a second-order phase transition, the phase transition is hysteresis-free. Near $T_c$, the up and down curves coincide, and the transition shows second-order phase transition behavior. At and above $T_c$, hysteresis losses become zero, and the magnetocaloric response M(h, T) becomes a truly reversible magnetocaloric effect that is the property of an efficient cooling cycle \cite{gutfleisch-2016,franco-2017}. 
	
	While the critical temperature $T_c$ increases with $r$, reflecting the enhanced stability of ordering in highly connected lattices, $\Delta S_M$ follows an inverse trend. The honeycomb lattice ($r=3$) achieves a higher $\Delta S_M$ compared to the square ($r=4$) and triangular lattice ($r=6$) configurations. This suggests that lower coordination facilitates a more rapid thermal randomization of spins near the transition point. The same pattern holds for the bilayer cases, as the introduction of an additional layer increases the coordination number or the connectedness, which enhances the stability of the ordering and decreases $\Delta S_M$. This again points out that lower coordination supports a more rapid thermal randomization of spins near the transition point. Furthermore, as shown in FIG. \ref{Hyst_img}, the hysteresis analysis indicates that the high entropy change for the low $r$ lattices shows a fast reduction in coercivity with the increase in temperature. Therefore, the lattices showing high $\Delta S_M$ show a fast reduction in the loop width per unit increase in temperature. Thus, the high entropy capacity of the low $r$ lattices is coupled with low coercivity.
	
	\section{Discussion and Outlook}
	\label{outlook}
	
	Here, the energy is in units of $J$, and the temperature in units of $(J/k_B)$. The magnetic field is shown in units of $g \mu_B h/ J \equiv h/J$(where, for simplification, we substitute $g=1$ and $\mu_B=1$). When we substitute $g=2$ and $ \mu_B = 5.788 \times 10^{-5} eV/\mathcal{T}$($\mathcal{T}$:Tesla) the value of the magnetic field becomes $\sim$ 2.6 $\mathcal{T}$ (assuming $J\approx 1 \; meV$).
	
	We observe a few discrepancies in low temperatures with the increase in the field strength. The scaling and universality are phenomena related to the critical region, which might follow outside the critical region \cite{franco-2006,smith-2014}. In our studies, we found that after a certain field value, particularly in the ferromagnetic region, the scaling and universality start to deviate from the master curve in the low temperature region. This is physically expected in view of the fact that no matter what the strength of the field is, the $S_M$ is always bound by the value of $k_Bln(2\sigma+1)$ per site, representing the saturation of the magnetization, denoting complete alignment of the magnetic moments \cite{smith-2014}. So, $-\Delta S_M$ will also be bound by a maximum value. The scaling relation given by Eq. \ref{n_T} has been derived from the Arrott-Noakes equation of state \cite{franco-2006}. The Arrott-Noakes equation of state \cite{arrott-1967} belongs to the group of models that are derived from the scaling relations and critical exponents, and written by applying the critical scaling for the entire temperature and field range. Therefore, the models do not consider the fact of saturation in magnetization. So, contrary to the true nature of the materials, Eq. \ref{n_T} presents $n$ as a field-independent quantity. But due to saturation in magnetization, after a certain field, the $-\Delta S_M$ and exponent $n$ do become field dependent \cite{smith-2014}. Our results also show the same behavior with more pronounced field dependence for lattices with low coordination numbers. 
	
	One of the important points is that the number of nearest neighbors or the coordination number ($r$) is also significant for the interpretation of the above results. The lattices with lower $r$ show more deviation than the lattices with higher $ r$. So, as magnetic interaction effectively depends on the number of nearest neighbors as well, let's say we define $r$-dependent magnetic interaction ($J$) as $J_{eff} \propto rJ$. Then the above concept (the deviation of $n$ from the $n\approx$ 1 to lower $n$ values for T<$T_c$ for the lattices) can also be seen as follows. For the six systems, the effective field $h/ J_{eff}$ will depend on the $r$ value and will not be the same for all. With the increase in the value of $h$, for lattices with low $r$, the $h/ J_{eff}$ will increase faster as compared to lattices with greater $r$. Finally, the pattern which is depicted by the results of the lattices with lower $r$ will be followed by the lattices with higher $r$ values, but at higher field ($h/ J_{eff}$) values. So, for the 2D systems, the magnetocaloric quantities follow the scaling laws up to a particular magnetic field($h/J$) value. At higher fields, some deviations from the power laws are expected and observed. 
	
\begin{figure}[H]
		\centering
		\includegraphics[width=16cm,height=6.5cm]{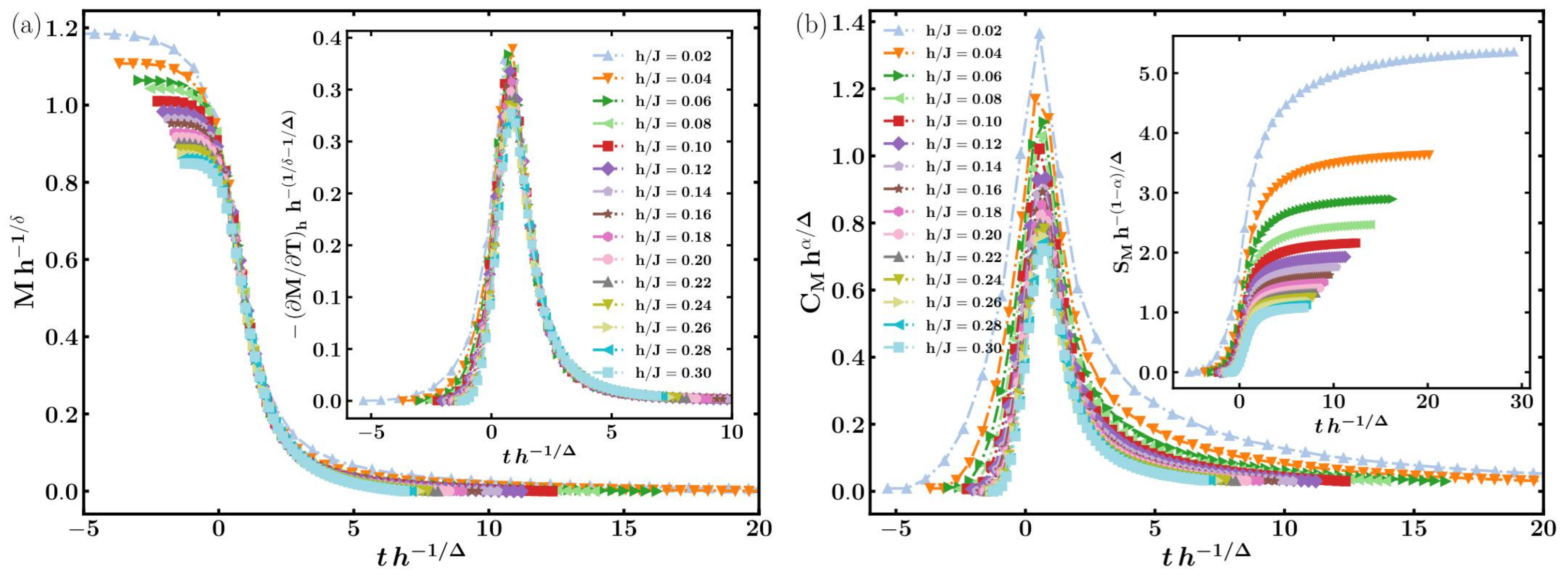}
		\caption{Scaling of (a) M and $\partial M/\partial T$ (in the inset); and (b) $C_M$ and $S_M$(in the inset) for honeycomb lattice as per the scaling laws}
		\label{Scaling_M+dmdt+s+c}
	\end{figure}

	In the results for the scaling and universality that are shown above we discussed the results only for the normalized $\Delta S_M/\Delta S_M^{Max}$, exponent $n$, $\Delta T_{ad}/\Delta T_{ad}^{Max}$ and $\Gamma_{M}/\Gamma_{M}^{Max}$ with respect to $\theta$. We also present the study of the scaling of these quantities with respect to $th^{-1/\Delta}$ as defined by their scaling relations discussed in section \ref{scaling_and_universality} and shown in FIG. \ref{dels_rcp_MCS}(c), \ref{dels_rcp_MCS}(e), \ref{delT_all_MCS}(b) and \ref{Gamma_all_MCS}(b), respectively. We observe some key differences in the results of these two kinds of scaling. At $\theta = 0$, the peak of these quantities reaches 1, but at $th^{-1/\Delta}$ = 0 the peak of these magnetocaloric quantities does not reach 1. In fact, we found that it actually depends on the coordination number of the lattice. The magnetocaloric quantities like $\Delta T_{ad}$ and exponent $n$ for which the peak values decrease with the increase in coordination number also show an increase when plotted as per their scaling relations. Similarly, the quantities like $\Delta S_M$ and $\Gamma_{M}$ show a decrease with the increase in the coordination number when plotted as per their scaling relations. Furthermore, the curves of the quantities $\Delta T_{ad}$ and $\Gamma_{M}$ when plotted as per their scaling relations do not collapse. As shown in FIG. \ref{Scaling_M+dmdt+s+c}, $M$ and $(\partial M /\partial T)_h$ show almost collapsing curves except at low temperatures but $C_M$ and $S_M$ curves does not collapse. This is because $C_M$ does not follow the scaling relations \cite{smith-2014}. Hence, the curves do not collapse. Similarly, the magnetic entropy $S_M$ (which is calculated from the $C_M$ as given by the relation \ref{eq_6}) when plotted as per the scaling relations does not show a collapse. Although the change in magnetic $\Delta S_M$ entropy shows a collapse following its scaling relation. 
	
	As discussed earlier, the coordination numbers for the honeycomb, square, and triangular lattices: (i) in the monolayer case are 3, 4, and 6; (ii) in the bilayer cases with open boundary conditions(along the $z$-axis only) are 4, 5, and 7; and (iii) in the bilayer cases with periodic boundary conditions are 5, 6, and 8. The results of the three lattices for the bilayer lattices with open boundary conditions are similar to their counterparts--the monolayer and bilayer lattices with periodic boundary conditions. For the bilayer lattices with open boundary conditions, the thermodynamic and magnetocaloric quantities, which are discussed as exhibiting increment/decrement as we go from monolayer to bilayer, the quantities show quantitatively intermediate kind of behavior for the particular lattice among the three cases. For example, $\Delta S_M$ of the bilayer cases without periodic boundary condition quantitatively lies in between $\Delta S_M$ of the monolayer lattice and the bilayer lattice with periodic boundary condition. In our investigations, we observe that the lattices with different names but the same coordination numbers have almost the same thermodynamic and magnetocaloric properties, which can be easily checked for the square bilayer and triangular monolayer cases in the results presented above. This is due to the topological invariance of the lattices. So, we can safely say that from the honeycomb lattice to the triangular bilayer lattice, we present a detailed investigation of the thermodynamic and magnetocaloric properties with varying coordination number from $r=3$ to $r=8$ for the 2D cases of the ferromagnetic Ising model. This is the main reason why the $J_{||}$ and $J_{\perp}$ values have been set constant throughout the study.
	
	We perform the Monte-Carlo studies of the Ising model for the six lattice structures considering the nearest-neighbor interactions. While considering the next nearest neighbor interaction in discussions of the model for the 2D square lattice, there are evidences of the presence of the strong first order, weak first order and second order transitions, in different paramagnetic regimes of the ratio of the first($J_{||}$) and second($J_2$) nearest neighbor interaction strengths ($\mathcal{R} = J_{||}/J_2$) \cite{gangat-2024}. The inclusion of the next-nearest neighbors creates a competition resulting in the formation of ferromagnetic, anti-ferromagnetic, super-anti-ferromagnetic, and paramagnetic phases \cite{gangat-2024,moran-lopez-1993}. Similarly, the universality class for the honeycomb and the triangular lattice with the next-nearest-neighbor interactions and different antiferromagnetic and ferromagnetic interactions for the different range of the ratio $\mathcal{R}$ is not defined clearly \cite{gessert-2026,malakis-2007} and requires a separate thorough investigation. Guerrero \textit{et al.} showed that just making $J_2$ as anti-ferromagnetic in the Ising model produces both normal and inverse magnetocaloric effect presented the existence of the normal and inverse magnetocaloric effect; existence of various phases; and different orders in the $J_1-J_2$ transverse Ising model with first and second nearest neighbor interactions using cluster mean field approach for ferromagnetic $J_{||}=1$ and anti-ferromagnetic $J_2=-1$ \cite{guerrero-2020}. This also produces different phases and varying orders of phase transitions. We can expect similar behavior in the bilayer cases of the other lattices. The effect of variation of $J_{\perp}$ from ferromagnetic to anti-ferromagnetic can also lead to first-order phase transitions \cite{abe-1970,chapman-2026}. Akinci \textit{et al.} found that with the increase in $\mathcal{R}$ the magnetic entropy change decreases for the bilayer square lattice using the mean field approximation\cite{aknc-2020}. Furthermore, they found that when the spin of one layer is kept constant at $\sigma=1/2$ and the $\sigma$ value of the second layer is increased from $1/2$ to $7/2$, the magnetic entropy increases, and it produces double $\Delta S_M$ peaks. Similar behavior can be expected from the bilayer honeycomb and triangular lattices. Considering the length of the paper, these questions will be examined in our future work.
	
	\section{Conclusion}
	\label{conc}
	We present a qualitative and quantitative investigation of the magnetocaloric properties that are most commonly quantified by the magnetic entropy change ($\Delta S_M$), adiabatic temperature change ($\Delta T_{ad}$), the magnetic Gr\"{u}neisen parameter ($\Gamma_{M}$), relative cooling power (RCP), and cooling capacity ($q$) and their power laws for two-dimensional (2D) monolayer and bilayer crystal structures that are the square, honeycomb, and triangular lattices by performing the Monte Carlo simulations of the Ising model. These crystal structures exhibit different magnetic responses because variations in lattice geometry alter the number of nearest neighbors, next-to-nearest neighbors, \textit{etc.,} thereby changing the number of magnetic exchange interactions between the magnetic dipoles. We study the Binder cumulant as a function of temperature to determine the precise transition temperature ($T_c$) of these systems. In both monolayer and bilayer cases, depending on the structure, we find that the critical temperature increases with the coordination number, yielding the lowest $T_c$ for the honeycomb lattice, which has the lowest coordination number, followed by the square and triangular lattices. On the contrary, we observed that $\Delta S_M$ is highest for the honeycomb lattice. We found that $\Delta S_M$ when normalized with its maximum value ($\Delta S_M^{Max}$), \textit{i.e.,} $\Delta S_M/\Delta S_M^{Max}$ as well as the exponent $n$ of $\Delta S_M$ portraying the field dependence of $\Delta S_M$ versus rescaled temperature ($\theta$) curves apparently collapse onto a single curve for each of the studied crystals in the range of low to moderate fields in the critical regime showing universality with respect to field, individually for all these structures. For all studied monolayer and bilayer crystal structures, the magnetic entropy change ($\Delta S_M$) curves surprisingly collapse onto a single universal profile up to a particular field, remaining invariant despite changes in coordination numbers that alter the effective magnetic exchange energy. Remarkably, the exponent $n$ versus $\theta$ also exhibits universal behavior at low fields for each of the lattices and especially for different lattices at low fields. This, in addition to showing the second-order nature of the phase transitions in the system, indicates that the phase transition is driven by the same phenomenological behavior. Contrary to the behavior of $\Delta S_M$, $\Delta T_{ad}$ is maximum for a triangular lattice having the highest coordination number among the three lattice structures and decreases as the coordination number decreases. However, like $-\Delta S^{}_M$, $\Gamma_{M}$ decreases with the coordination number, reaching its maximum for the honeycomb lattice. Like $\Delta S_M$, $\Delta T_{ad}$ and $\Gamma_{M}$ also show scaling behavior. Although $\Delta S_M$ peak values decrease with the increase in the coordination number, the quantities RCP and $q$ do not show much variation because the width of the $\Delta S_M$ curves at half maximum increases. In the study of the scaling laws for $\Delta S_M$, RCP, and $q$, we found that these properties follow power laws for the conventional magnetic-field range of 2.6 Tesla (assuming $J\approx 1 \; meV$). Finally, the hysteresis analysis shows that lattices with a low number of nearest neighbors exhibit high $\Delta S_M$ and a rapid reduction in coercivity with increasing temperature.
	
	\section{ACKNOWLEDGMENT}
	The authors thank the technical and software support of the High-Performance Computing (HPC) lab at VIT-AP University for providing computational resources. K.S. acknowledges the support of the VIT-AP University Research Grant in Engineering Management and Sciences (RGEMS) grant with order number VIT-AP/SpoRIC/RGEMS/2024-2025/005. The authors acknowledge the insightful discussions with Arghya Taraphder.
	
	\bibliographystyle{elsarticle-num}
	\bibliography{cas-refs2}

\begin{thebibliography}{10}
\expandafter\ifx\csname url\endcsname\relax
  \def\url#1{\texttt{#1}}\fi
\expandafter\ifx\csname urlprefix\endcsname\relax\def\urlprefix{URL }\fi
\expandafter\ifx\csname href\endcsname\relax
  \def\href#1#2{#2} \def\path#1{#1}\fi

\bibitem{weiss-1917}
P.~Weiss, A.~Piccard, {Le phénomène magnétocalorique}, Journal de Physique
  Théorique et Appliquée 7~(1) (1917) 103--109.
\newblock \href {https://doi.org/10.1051/jphystap:019170070010300}
  {\path{doi:10.1051/jphystap:019170070010300}}.

\bibitem{tishin-2010}
A.~Tishin, Y.~Spichkin, {The Magnetocaloric Effect and its Applications}, 2010.
\newblock \href {https://doi.org/10.1201/9781420033373}
  {\path{doi:10.1201/9781420033373}}.

\bibitem{mcnerney-2025}
G.~W. McNerney, {The environmental impact and energy efficiency of
  magnetocaloric refrigeration systems}, International Journal of Environmental
  Sustainability and Green Technologies 15~(1) (2025) 1--13.
\newblock \href {https://doi.org/10.4018/ijesgt.369160}
  {\path{doi:10.4018/ijesgt.369160}}.

\bibitem{lucia-2024}
U.~Lucia, G.~Grisolia, {Magnetocaloric refrigeration in the context of
  sustainability: A review of thermodynamic bases, the state of the art, and
  future prospects}, Energies 17~(14) (2024) 3585.
\newblock \href {https://doi.org/10.3390/en17143585}
  {\path{doi:10.3390/en17143585}}.

\bibitem{gschneidner-2008}
K.~Gschneidner, V.~Pecharsky, {Thirty years of near room temperature magnetic
  cooling: Where we are today and future prospects}, International Journal of
  Refrigeration 31~(6) (2008) 945--961.
\newblock \href {https://doi.org/10.1016/j.ijrefrig.2008.01.004}
  {\path{doi:10.1016/j.ijrefrig.2008.01.004}}.

\bibitem{pecharsky-1997}
V.~K. Pecharsky, K.~A. Gschneidner, Jr, {Giant Magnetocaloric Effect
  InGD5(SI2GE2)}, Physical Review Letters 78~(23) (1997) 4494--4497.
\newblock \href {https://doi.org/10.1103/physrevlett.78.4494}
  {\path{doi:10.1103/physrevlett.78.4494}}.

\bibitem{debye-1926}
P.~Debye, {Einige Bemerkungen zur Magnetisierung bei tiefer Temperatur},
  Annalen der Physik 386~(25) (1926) 1154--1160.
\newblock \href {https://doi.org/10.1002/andp.19263862517}
  {\path{doi:10.1002/andp.19263862517}}.

\bibitem{giauque-1927}
W.~F. Giauque, {A Thermodynamic Treatment of Certain Magnetic Effects. A
  Proposed Method of Producing Temperatures considerably Below 1° Absolute},
  Journal of the American Chemical Society 49~(8) (1927) 1864--1870.
\newblock \href {https://doi.org/10.1021/ja01407a003}
  {\path{doi:10.1021/ja01407a003}}.

\bibitem{franco-2017}
V.~Franco, J.~Blázquez, J.~Ipus, J.~Law, L.~Moreno-Ramírez, A.~Conde,
  {Magnetocaloric effect: From materials research to refrigeration devices},
  Progress in Materials Science 93 (2017) 112--232.
\newblock \href {https://doi.org/10.1016/j.pmatsci.2017.10.005}
  {\path{doi:10.1016/j.pmatsci.2017.10.005}}.

\bibitem{gutfleisch-2016}
O.~Gutfleisch, T.~Gottschall, M.~Fries, D.~Benke, I.~Radulov, K.~P. Skokov,
  H.~Wende, M.~Gruner, M.~Acet, P.~Entel, M.~Farle, {Mastering hysteresis in
  magnetocaloric materials}, Philosophical Transactions of the Royal Society A
  Mathematical Physical and Engineering Sciences 374~(2074) (2016) 20150308.
\newblock \href {https://doi.org/10.1098/rsta.2015.0308}
  {\path{doi:10.1098/rsta.2015.0308}}.

\bibitem{widom-1965}
B.~Widom, {Equation of state in the neighborhood of the critical point}, The
  Journal of Chemical Physics 43~(11) (1965) 3898--3905.
\newblock \href {https://doi.org/10.1063/1.1696618}
  {\path{doi:10.1063/1.1696618}}.

\bibitem{oesterreicher-1984}
H.~Oesterreicher, F.~T. Parker, {Magnetic cooling near Curie temperatures above
  300 K}, Journal of Applied Physics 55~(12) (1984) 4334--4338.
\newblock \href {https://doi.org/10.1063/1.333046}
  {\path{doi:10.1063/1.333046}}.

\bibitem{franco-2006}
V.~Franco, J.~S. Blázquez, A.~Conde, {Field dependence of the magnetocaloric
  effect in materials with a second order phase transition: A master curve for
  the magnetic entropy change}, Applied Physics Letters 89~(22) (11 2006).
\newblock \href {https://doi.org/10.1063/1.2399361}
  {\path{doi:10.1063/1.2399361}}.

\bibitem{arrott-1967}
A.~Arrott, J.~E. Noakes, {Approximate Equation of State For Nickel Near its
  Critical Temperature}, Physical Review Letters 19~(14) (1967) 786--789.
\newblock \href {https://doi.org/10.1103/physrevlett.19.786}
  {\path{doi:10.1103/physrevlett.19.786}}.

\bibitem{lyubina-2011}
J.~Lyubina, M.~D. Kuz’min, K.~Nenkov, O.~Gutfleisch, M.~Richter, D.~L.
  Schlagel, T.~A. Lograsso, K.~A. Gschneidner, {Magnetic field dependence of
  the maximum magnetic entropy change}, Physical Review B 83~(1) (1 2011).
\newblock \href {https://doi.org/10.1103/physrevb.83.012403}
  {\path{doi:10.1103/physrevb.83.012403}}.

\bibitem{liu-2026}
X.~Liu, E.~Lv, X.~Cui, H.~Ge, F.~Song, Z.~Tian, G.~Su, K.~Zhao, J.~Xiang,
  P.~Sun, W.~Li, {ISing Supercriticality and Universal Magnetocalorics in
  Spiral Antiferromagnet ND 3 BWO 9}, Physical Review Letters 136~(24) (5
  2026).
\newblock \href {https://doi.org/10.1103/nqcw-pz8v}
  {\path{doi:10.1103/nqcw-pz8v}}.

\bibitem{franco-2008}
V.~Franco, A.~Conde, J.~M. Romero-Enrique, J.~S. Blázquez, {A universal curve
  for the magnetocaloric effect: an analysis based on scaling relations},
  Journal of Physics Condensed Matter 20~(28) (2008) 285207.
\newblock \href {https://doi.org/10.1088/0953-8984/20/28/285207}
  {\path{doi:10.1088/0953-8984/20/28/285207}}.

\bibitem{guerrero-2020}
A.~I. Guerrero, S.~G. J, {Magnetocaloric effect in the $J_1–J_2$ transverse
  Ising model}, Journal of Magnetism and Magnetic Materials 514 (2020) 167140.
\newblock \href {https://doi.org/10.1016/j.jmmm.2020.167140}
  {\path{doi:10.1016/j.jmmm.2020.167140}}.

\bibitem{alzate-cardona-2019}
J.~D. Alzate-Cardona, J.~S. Salcedo-Gallo, D.~F. Rodríguez-Patiño, C.~D.
  Acosta-Medina, E.~Restrepo-Parra, {Unveiling a scaling and universal behavior
  for the magnetocaloric effect in cubic crystal structures: a Monte Carlo
  simulation}, Scientific Reports 9~(1) (2019) 5228.
\newblock \href {https://doi.org/10.1038/s41598-019-41321-y}
  {\path{doi:10.1038/s41598-019-41321-y}}.

\bibitem{patra-2024}
L.~Patra, Y.~Quan, B.~Liao, {Impact of dimensionality on the magnetocaloric
  effect in two-dimensional magnets}, Journal of Applied Physics 136~(2) (7
  2024).
\newblock \href {https://doi.org/10.1063/5.0218007}
  {\path{doi:10.1063/5.0218007}}.

\bibitem{schmidt-2007}
B.~Schmidt, P.~Thalmeier, N.~Shannon, {Magnetocaloric effect in the frustrated
  square lattice $J_1 - J_2$ model}, Physical Review B 76~(12) (9 2007).
\newblock \href {https://doi.org/10.1103/physrevb.76.125113}
  {\path{doi:10.1103/physrevb.76.125113}}.

\bibitem{zeng-2012}
R.~Zeng, S.~Q. Wang, G.~D. Du, J.~L. Wang, J.~C. Debnath, P.~Shamba, Z.~Y.
  Fang, S.~X. Dou, {Abnormal magnetic behaviors and large magnetocaloric effect
  in MnPS3 nanoparticles}, Journal of Applied Physics 111~(7) (3 2012).
\newblock \href {https://doi.org/10.1063/1.3679409}
  {\path{doi:10.1063/1.3679409}}.

\bibitem{hou-2018}
Y.~S. Hou, J.~H. Yang, H.~J. Xiang, X.~G. Gong, {First-principles study of the
  magnetic interactions in honeycomb Na2IrO3}, Physical review. B./Physical
  review. B 98~(9) (9 2018).
\newblock \href {https://doi.org/10.1103/physrevb.98.094401}
  {\path{doi:10.1103/physrevb.98.094401}}.

\bibitem{gao-2022}
F.~Gao, J.~Sheng, W.~Ren, Q.~Zhang, X.~Luo, J.~Qi, M.~Cong, B.~Li, L.~Wu,
  Z.~Zhang, {Incommensurate spin density wave and magnetocaloric effect in the
  metallic triangular lattice HoAl2Ge2}, Physical review. B./Physical review. B
  106~(13) (10 2022).
\newblock \href {https://doi.org/10.1103/physrevb.106.134426}
  {\path{doi:10.1103/physrevb.106.134426}}.

\bibitem{gong-2023}
J.~Gong, L.~Tian, L.~Zhang, Z.~Mo, Y.~Wang, J.~Shen, {Magnetism and cryogenic
  magnetocaloric effect of triangular-lattice LnOF (Ln = Gd, Dy, Ho, and Er)
  compounds}, Journal of Rare Earths 43~(1) (2023) 98--104.
\newblock \href {https://doi.org/10.1016/j.jre.2023.10.005}
  {\path{doi:10.1016/j.jre.2023.10.005}}.

\bibitem{nobrega-2006_2}
E.~P. Nóbrega, N.~A. De~Oliveira, P.~J. Von~Ranke, A.~Troper, {The
  magnetocaloric effect in R5Si4(R = Gd, Tb): a Monte Carlo calculation},
  Journal of Physics Condensed Matter 18~(4) (2006) 1275--1283.
\newblock \href {https://doi.org/10.1088/0953-8984/18/4/013}
  {\path{doi:10.1088/0953-8984/18/4/013}}.

\bibitem{li-2026}
W.~Li, Z.~Liu, D.~Xie, {Magnetocaloric properties and critical behavior of
  rare-earth perovskite LaMnO3}, Journal of Magnetism and Magnetic Materials
  655 (2026) 174333.
\newblock \href {https://doi.org/10.1016/j.jmmm.2026.174333}
  {\path{doi:10.1016/j.jmmm.2026.174333}}.

\bibitem{masrour-2019}
R.~Masrour, A.~Jabar, M.~S.~B. Kraiem, M.~Ellouze, N.~Randrianantoandro,
  S.~Labidi, Experimental and monte carlo simulation studies of the
  magnetocaloric effect in $r_{2}fe_{17}$ compounds ($r = $nd, gd), Indian
  Journal of Physics 94~(11) (2019) 1717--1724.
\newblock \href {https://doi.org/10.1007/s12648-019-01615-3}
  {\path{doi:10.1007/s12648-019-01615-3}}.

\bibitem{iqbal-2026-mft}
B.~Iqbal, K.~Sarkar, {Mean Field Analysis of Magnetocaloric Effect of the 2D
  Ferromagnetic Ising Model (In progress)}.

\bibitem{landau-2015}
D.~P. Landau, K.~Binder, {A guide to Monte Carlo simulations in Statistical
  Physics}, Cambridge University Press, 2015.

\bibitem{metropolis-1953}
N.~Metropolis, A.~W. Rosenbluth, M.~N. Rosenbluth, A.~H. Teller, E.~Teller,
  {Equation of state calculations by fast computing machines}, The Journal of
  Chemical Physics 21~(6) (1953) 1087--1092.
\newblock \href {https://doi.org/10.1063/1.1699114}
  {\path{doi:10.1063/1.1699114}}.

\bibitem{glauber-1963}
R.~J. Glauber, {Time-Dependent Statistics of the Ising model}, Journal of
  Mathematical Physics 4~(2) (1963) 294--307.
\newblock \href {https://doi.org/10.1063/1.1703954}
  {\path{doi:10.1063/1.1703954}}.

\bibitem{zia-2007}
R.~K.~P. Zia, B.~Schmittmann, {Probability currents as principal
  characteristics in the statistical mechanics of non-equilibrium steady
  states}, Journal of Statistical Mechanics Theory and Experiment 2007~(07)
  (2007) P07012.
\newblock \href {https://doi.org/10.1088/1742-5468/2007/07/p07012}
  {\path{doi:10.1088/1742-5468/2007/07/p07012}}.

\bibitem{wolff-1989}
U.~Wolff, {Collective Monte Carlo updating for spin systems}, Physical Review
  Letters 62~(4) (1989) 361--364.
\newblock \href {https://doi.org/10.1103/physrevlett.62.361}
  {\path{doi:10.1103/physrevlett.62.361}}.

\bibitem{luijten-2007}
E.~Luijten, {Introduction to Cluster Monte Carlo Algorithms}, 2007.
\newblock \href {https://doi.org/10.1007/3-540-35273-2_1}
  {\path{doi:10.1007/3-540-35273-2_1}}.

\bibitem{wakabayashi-1999}
K.~Wakabayashi, M.~Fujita, H.~Ajiki, M.~Sigrist, {Electronic and magnetic
  properties of nanographite ribbons}, Physical review. B, Condensed matter
  59~(12) (1999) 8271--8282.
\newblock \href {https://doi.org/10.1103/physrevb.59.8271}
  {\path{doi:10.1103/physrevb.59.8271}}.

\bibitem{morita-2015}
S.~Morita, S.~Suzuki, {Phase Transition of Two-Dimensional Ising Models on the
  Honeycomb and Related Lattices with Striped Random Impurities}, Journal of
  Statistical Physics 162~(1) (2015) 123--138.
\newblock \href {https://doi.org/10.1007/s10955-015-1400-0}
  {\path{doi:10.1007/s10955-015-1400-0}}.

\bibitem{nobrega-2006}
E.~P. Nobrega, N.~A. De~Oliveira, P.~J. Von~Ranke, A.~Troper, {Magnetocaloric
  effect in $(Gd_xTb_{1-x})_5 Si_4$ by Monte Carlo simulations}, Physical
  Review B 74~(14) (10 2006).
\newblock \href {https://doi.org/10.1103/physrevb.74.144429}
  {\path{doi:10.1103/physrevb.74.144429}}.

\bibitem{de-castro-2025}
L.~De~Castro, C.~Morais, F.~Zimmer, M.~Schmidt, {Size effects on the magnetism
  and magnetocaloric properties of a polymeric Ising chain}, Physica E
  Low-dimensional Systems and Nanostructures (2025) 116272\href
  {https://doi.org/10.1016/j.physe.2025.116272}
  {\path{doi:10.1016/j.physe.2025.116272}}.

\bibitem{li-2023}
B.-C. Li, W.~Wang, {Influence of a new long-range interaction on the magnetic
  properties of a 2D Ising layered model by using Monte Carlo method}, Chinese
  Journal of Physics 87 (2023) 525--539.
\newblock \href {https://doi.org/10.1016/j.cjph.2023.12.023}
  {\path{doi:10.1016/j.cjph.2023.12.023}}.

\bibitem{yang-2022}
M.~Yang, F.~Wang, W.~Wang, B.-C. Li, J.-Q. Lv, {Insight into magnetic
  properties and magnetocaloric effect of an Ising-type polyhedral chain},
  Polymer 246 (2022) 124756.
\newblock \href {https://doi.org/10.1016/j.polymer.2022.124756}
  {\path{doi:10.1016/j.polymer.2022.124756}}.

\bibitem{gruneisen-1908}
E.~Grüneisen, {Über die thermische Ausdehnung und die spezifische Wärme der
  Metalle}, Annalen der Physik 331~(6) (1908) 211--216.
\newblock \href {https://doi.org/10.1002/andp.19083310611}
  {\path{doi:10.1002/andp.19083310611}}.

\bibitem{gruneisen-1912}
E.~Grüneisen, {Theorie des festen Zustandes einatomiger Elemente}, Annalen der
  Physik 344~(12) (1912) 257--306.
\newblock \href {https://doi.org/10.1002/andp.19123441202}
  {\path{doi:10.1002/andp.19123441202}}.

\bibitem{yu-2020}
Y.~Yu, R.~Liu, L.~Zhang, G.~Yao, Q.~Wang, J.~Zhu, S.~Yang, W.~Cui, {Enhanced
  magnetocaloric effects in Gd$_2$In$_{1-x}$Al$_x$ (0.4 $\leq$ x $\leq$ 1)
  system by the hysteresis-free metamagnetism}, Journal of Magnetism and
  Magnetic Materials 524 (2020) 167648.
\newblock \href {https://doi.org/10.1016/j.jmmm.2020.167648}
  {\path{doi:10.1016/j.jmmm.2020.167648}}.

\bibitem{zhu-2003}
L.~Zhu, M.~Garst, A.~Rosch, Q.~Si, {Universally diverging Grüneisen parameter
  and the magnetocaloric effect close to quantum critical points}, Physical
  Review Letters 91~(6) (2003) 066404.
\newblock \href {https://doi.org/10.1103/physrevlett.91.066404}
  {\path{doi:10.1103/physrevlett.91.066404}}.

\bibitem{stanley-1999}
H.~E. Stanley, {Scaling, universality, and renormalization: Three pillars of
  modern critical phenomena}, Reviews of Modern Physics 71~(2) (1999)
  S358--S366.
\newblock \href {https://doi.org/10.1103/revmodphys.71.s358}
  {\path{doi:10.1103/revmodphys.71.s358}}.

\bibitem{smith-2014}
A.~Smith, K.~K. Nielsen, C.~R.~H. Bahl, {Scaling and universality in
  magnetocaloric materials}, Physical Review B 90~(10) (9 2014).
\newblock \href {https://doi.org/10.1103/physrevb.90.104422}
  {\path{doi:10.1103/physrevb.90.104422}}.

\bibitem{liu-2019}
W.~Liu, Y.~Wang, J.~Fan, L.~Pi, M.~Ge, L.~Zhang, Y.~Zhang, {Field-dependent
  anisotropic magnetic coupling in layered ferromagnetic Fe3-xGeTe2}, Physical
  review. B./Physical review. B 100~(10) (9 2019).
\newblock \href {https://doi.org/10.1103/physrevb.100.104403}
  {\path{doi:10.1103/physrevb.100.104403}}.

\bibitem{franco-2010}
V.~Franco, A.~Conde, D.~Sidhaye, B.~L.~V. Prasad, P.~Poddar, S.~Srinath, M.~H.
  Phan, H.~Srikanth, {Field dependence of the magnetocaloric effect in
  core-shell nanoparticles}, Journal of Applied Physics 107~(9) (4 2010).
\newblock \href {https://doi.org/10.1063/1.3335514}
  {\path{doi:10.1063/1.3335514}}.

\bibitem{bonilla-2010}
C.~M. Bonilla, J.~Herrero-Albillos, F.~Bartolomé, L.~M. García,
  M.~Parra-Borderías, V.~Franco, {Universal behavior for magnetic entropy
  change in magnetocaloric materials: An analysis on the nature of phase
  transitions}, Physical Review B 81~(22) (6 2010).
\newblock \href {https://doi.org/10.1103/physrevb.81.224424}
  {\path{doi:10.1103/physrevb.81.224424}}.

\bibitem{iqbal-2024}
B.~Iqbal, K.~Sarkar, {Monte Carlo Simulation of An-Isotropic Ising model using
  Metropolis and Wolff algorithm}, Springer proceedings in physics (2024)
  9--28\href {https://doi.org/10.1007/978-981-96-0828-7_2}
  {\path{doi:10.1007/978-981-96-0828-7_2}}.

\bibitem{malakis-2014}
A.~Malakis, N.~G. Fytas, G.~G\"ulpinar, Critical binder cumulant and
  universality: Fortuin-kasteleyn clusters and order-parameter fluctuations,
  Phys. Rev. E 89 (2014) 042103.
\newblock \href {https://doi.org/10.1103/PhysRevE.89.042103}
  {\path{doi:10.1103/PhysRevE.89.042103}}.

\bibitem{kamieniarz-1993}
G.~Kamieniarz, H.~W.~J. Blote, {Universal ratio of magnetization moments in
  two-dimensional Ising models}, Journal of Physics A Mathematical and General
  26~(2) (1993) 201--212.
\newblock \href {https://doi.org/10.1088/0305-4470/26/2/009}
  {\path{doi:10.1088/0305-4470/26/2/009}}.

\bibitem{yamamoto-2009}
D.~Yamamoto, {Correlated cluster mean-field theory for spin systems}, Physical
  Review B 79~(14) (4 2009).
\newblock \href {https://doi.org/10.1103/physrevb.79.144427}
  {\path{doi:10.1103/physrevb.79.144427}}.

\bibitem{spaldin-2010}
N.~A. Spaldin, \href{https://doi.org/10.1017/cbo9780511781599}{{Magnetic
  materials}}, 2010.
\newblock \href {https://doi.org/10.1017/cbo9780511781599}
  {\path{doi:10.1017/cbo9780511781599}}.
\newline\urlprefix\url{https://doi.org/10.1017/cbo9780511781599}

\bibitem{yin-2009}
J.~Yin, D.~P. Landau, {Phase diagram and critical behavior of the
  square-lattice Ising model with competing nearest-neighbor and
  next-nearest-neighbor interactions}, Physical Review E 80~(5) (2009) 051117.
\newblock \href {https://doi.org/10.1103/physreve.80.051117}
  {\path{doi:10.1103/physreve.80.051117}}.

\bibitem{gangat-2024}
A.~A. Gangat, {Weak first-order phase transitions in the frustrated square
  lattice J1-J2 classical Ising model}, Physical review. B./Physical review. B
  109~(10) (3 2024).
\newblock \href {https://doi.org/10.1103/physrevb.109.104419}
  {\path{doi:10.1103/physrevb.109.104419}}.

\bibitem{moran-lopez-1993}
J.~L. Morán-López, F.~Aguilera-Granja, J.~M. Sanchez, {First-order phase
  transitions in the Ising square lattice with first- and second-neighbor
  interactions}, Physical review. B, Condensed matter 48~(5) (1993) 3519--3522.
\newblock \href {https://doi.org/10.1103/physrevb.48.3519}
  {\path{doi:10.1103/physrevb.48.3519}}.

\bibitem{gessert-2026}
D.~Gessert, M.~Weigel, W.~Janke, {Frustrated Ising model on the honeycomb
  lattice: Metastability and universality}, Physical review. B./Physical
  review. B 113~(10) (2 2026).
\newblock \href {https://doi.org/10.1103/gqwy-n84r}
  {\path{doi:10.1103/gqwy-n84r}}.

\bibitem{malakis-2007}
A.~Malakis, N.~Fytas, P.~Kalozoumis, {First-order transition features of the
  triangular Ising model with nearest- and next-nearest-neighbor
  antiferromagnetic interactions}, Physica A Statistical Mechanics and its
  Applications 383~(2) (2007) 351--371.
\newblock \href {https://doi.org/10.1016/j.physa.2007.04.051}
  {\path{doi:10.1016/j.physa.2007.04.051}}.

\bibitem{abe-1970}
R.~Abe, {Some Remarks on Perturbation Theory and Phase Transition with an
  Application to Anisotropic Ising Model}, Progress of Theoretical Physics
  44~(2) (1970) 339--347.
\newblock \href {https://doi.org/10.1143/ptp.44.339}
  {\path{doi:10.1143/ptp.44.339}}.

\bibitem{chapman-2026}
J.~Chapman, J.~Gidziunas, B.~Tomasello, S.~Carr, {The Widom line in the Ising
  model on a decorated bilayer lattice}, arXiv (Cornell University) (4 2026).
\newblock \href {https://doi.org/10.48550/arxiv.2604.11606}
  {\path{doi:10.48550/arxiv.2604.11606}}.

\bibitem{aknc-2020}
{\"U}.~Ak{\i}nc{\i}, Y.~Y{\"u}ksel, {Magnetocaloric effect in bilayer material:
  Spin-S Ising model}, arXiv (Cornell University) (12 2020).
\newblock \href {https://doi.org/10.48550/arxiv.2012.02161}
  {\path{doi:10.48550/arxiv.2012.02161}}.

\end{thebibliography}
	\appendix
	\section{Finite Size scaling analysis}
	\label{appdx_FSS}
	The finite-size effect of the magnetocaloric properties$\Delta S_M$ and $\Delta T_{ad}$. The magnetocaloric properties show some differences at low fields up to $h/J=0.08$ and up to a lattice size of $\mathscr{L}=16$. After that, the curves start collapsing.
	\begin{figure}[H]
		\centering
		\includegraphics[width=16cm,height=18cm]{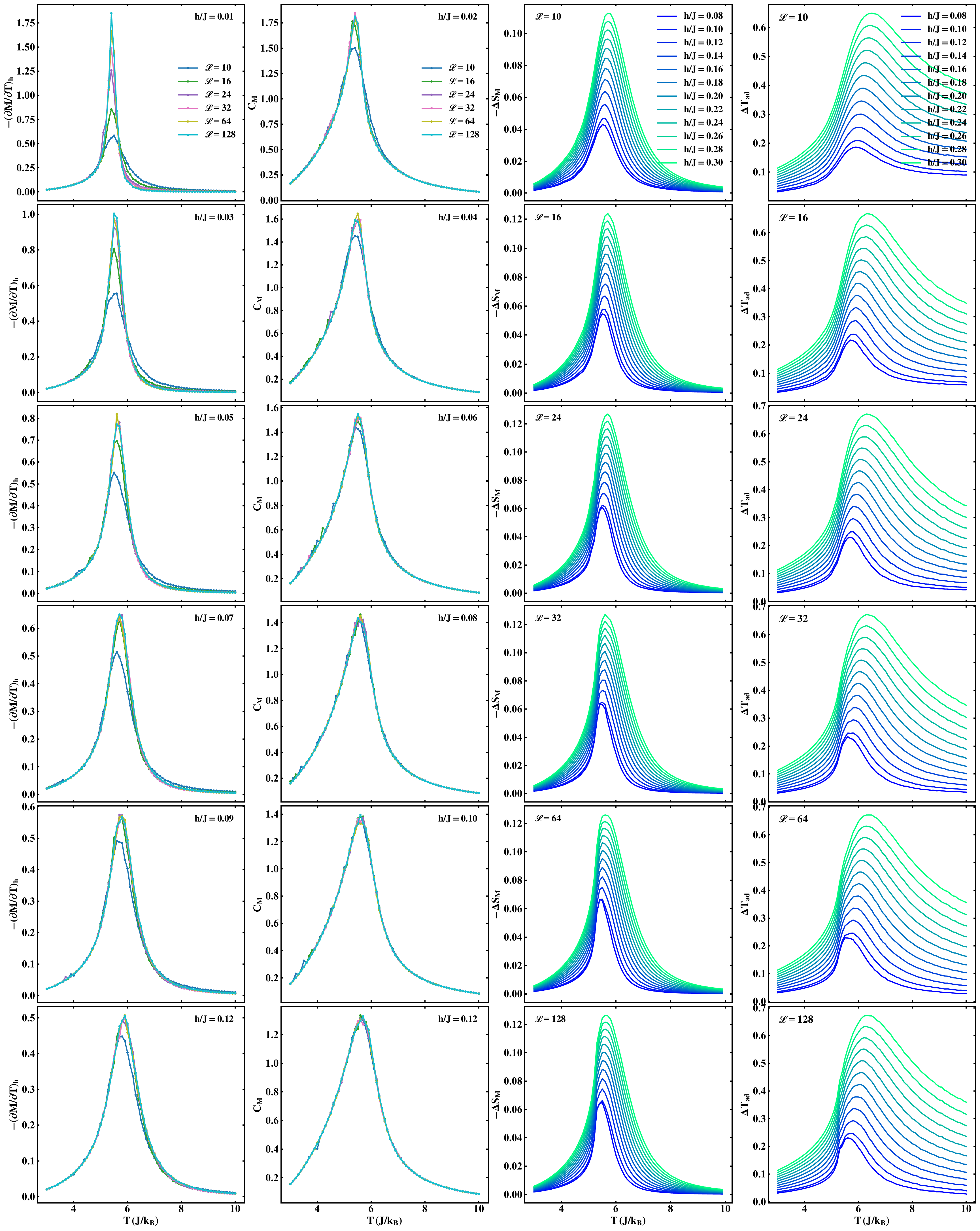}
		\caption{Variation of (fist column) $(\frac{\partial M}{\partial T})_h$ with T (second column) $C_M$ with T, (third column) $\Delta S_M$ with T and (fourth column) $\Delta T_{ad}$ with T of bilayer triangular lattice for different lattice sizes at different field values.}
		\label{FSS_VLS}
	\end{figure}
\end{document}